\documentclass{article}

\usepackage{PRIMEarxiv}

\usepackage{graphicx}
\usepackage[section]{algorithm}
\usepackage{algorithmic}
\usepackage{threeparttable}
\usepackage{amsthm}
\usepackage{amsmath,amssymb,amsfonts}
\usepackage{bm}
\usepackage{color}
\usepackage{ragged2e}
\usepackage{subfigure}
\usepackage{mathrsfs}
\usepackage{enumitem}
\usepackage{array}
\usepackage{url}
\usepackage{multirow}
\usepackage{booktabs}
\usepackage[T1]{fontenc}

\newcommand{\PreserveBackslash}[1]{\let\temp=\\#1\let\\=\temp}
\newcolumntype{C}[1]{>{\PreserveBackslash\centering}p{#1}}
\newcolumntype{R}[1]{>{\PreserveBackslash\raggedleft}p{#1}}
\newcolumntype{L}[1]{>{\PreserveBackslash\raggedright}p{#1}}

\renewcommand{\raggedright}{\leftskip=0pt \rightskip=0pt plus 0cm}

\AtBeginDocument{
  \providecommand\BibTeX{{
    \normalfont B\kern-0.5em{\scshape i\kern-0.25em b}\kern-0.8em\TeX}}}

\graphicspath{{media/}}     

\title{Blockchain-empowered Federated Learning: Benefits, Challenges, and Solutions
}

\author{
  Zeju~Cai, Jianguo~Chen, Yuting Fan, Zibin Zheng \\
  School of Software Engineering \\
  Sun Yat-sen University \\
  Zhuhai, Guangdong, P.R China \\
  \And
  Keqin~Li \\
  Department of Computer Science \\
  State University of New York \\
  New Paltz, NY, USA \\
}

\begin{document}
\include{multicol}
\maketitle

\begin{abstract}
  Federated learning (FL) is a distributed machine learning approach that protects user data privacy by training models locally on clients and aggregating them on a parameter server.
  While effective at preserving privacy, FL systems face limitations such as single points of failure, lack of incentives, and inadequate security.
  To address these challenges, blockchain technology is integrated into FL systems to provide stronger security, fairness, and scalability.
  However, blockchain-empowered FL (BC-FL) systems introduce additional demands on network, computing, and storage resources.
  This survey provides a comprehensive review of recent research on BC-FL systems, analyzing the benefits and challenges associated with blockchain integration.
  We explore why blockchain is applicable to FL, how it can be implemented, and the challenges and existing solutions for its integration.
  Additionally, we offer insights on future research directions for the BC-FL system.

\end{abstract}


\section{Introduction}
Artificial Intelligence (AI) technologies drive the Fourth Industrial Revolution, with user data being essential for training diverse Machine Learning (ML) models \cite{meurisch2021data}.
Training high-quality ML models often involves a centralized approach, necessitating internal storage of user data.
This raises privacy concerns \cite{bashir2015online,nadikattu2018iot,isaak2018user} and highlights the need for stringent privacy protections \cite{yin2021comprehensive}.
In recent years, regions such as the European Union \cite{tikkinen2018eu,mondschein2019eu}, the United States \cite{boyne2018data}, and Singapore \cite{yin2021comprehensive} have enacted relevant laws and regulations to govern the use of personal data, enhancing privacy protection but potentially hindering the utilization of high-quality data.

Federated Learning (FL) is a privacy-preserving distributed machine learning paradigm that balances user data protection and effective utilization \cite{mcmahan2017communication,wang2020federated,liu2020secure_5g}.
FL involves training local models on user devices and aggregating these local models into a global model on a server without requiring users to upload their data, addressing the aforementioned privacy concerns.
Initially applied to training Gboard \cite{ding2020incentive}, FL has proven successful.
Its potential extends beyond this, as it can also address the issue of data silos.
Data silos refer to the isolated or dispersed nature of data, making access to this data extremely challenging \cite{li2022federated}.
One cause of data silos is the reluctance of organizations to share data due to privacy or competitive concerns.
For instance, due to privacy protection, hospitals may be unwilling to share patient data \cite{nguyen2022federated}.
In summary, the judicious use of FL can break down data barriers, leading to its widespread application in healthcare \cite{li2023review,rani2023federated}, finance \cite{zheng2021federated,zhang2023federated}, industry \cite{arunan2023federated,zeng2023hfedms} and so on.

While privacy protection and data utilization benefits have popularized FL across industry and academia, they also introduce specific challenges.
First, there is a lack of trust among nodes within the FL system \cite{ji2023lafed,yang2024explainable}.
Nodes may worry that their training contributions will be intentionally tampered with or miscalculated, damaging their reputation and deserved rewards.
Second, FL systems are vulnerable to attacks from malicious nodes \cite{abou2023mitfed,kalapaaking2023blockchain}.
Malicious users may intentionally provide incorrect information to prevent model convergence and disrupt model training, while malicious servers can recover users' training data from the uploaded models.
Third, FL is prone to single point of failure issues \cite{wang2022blockchain2}.
In traditional FL architectures, the central server is responsible for aggregating and updating global model parameters.
If the central server is attacked or fails, the entire system's operation is severely affected, leading to interruptions in the training process, data loss, and irrecoverable model states.

Blockchain is essentially a distributed ledger, and its successful application in cryptocurrencies demonstrates its potential to build trust, security, and transparency \cite{xu2023survey,monrat2019survey,dai2019blockchain}.
Consequently, numerous studies have integrated blockchain with FL systems to enhance functionality, creating blockchain-empowered FL (BC-FL) systems.
Analyzing existing BC-FL literature, we find that blockchain's enhancement of different aspects of FL originates from its distinct properties.
First, blockchain's transparency and immutability can alleviate the lack of trust among nodes within the FL system.
By recording data requiring consensus in the FL system on the blockchain, these data cannot be tampered with by malicious nodes, enhancing trust relationships.
Second, through cross-validation of blockchain nodes and other mechanisms, the resistance of the FL system to malicious nodes is improved.
Finally, blockchain can replace the centralized server to avoid single point of failure issues.
By designing a reasonable consensus mechanism, suitable clients can be selected to undertake model aggregation tasks in each communication round.
With the advent of blockchain 2.0, users can develop smart contracts running on the blockchain, endowing BC-FL with greater scalability for automatically running various algorithms \cite{aggarwal2021blockchain}.

\begin{figure*}
    \centering
    \includegraphics[width=4.5in]{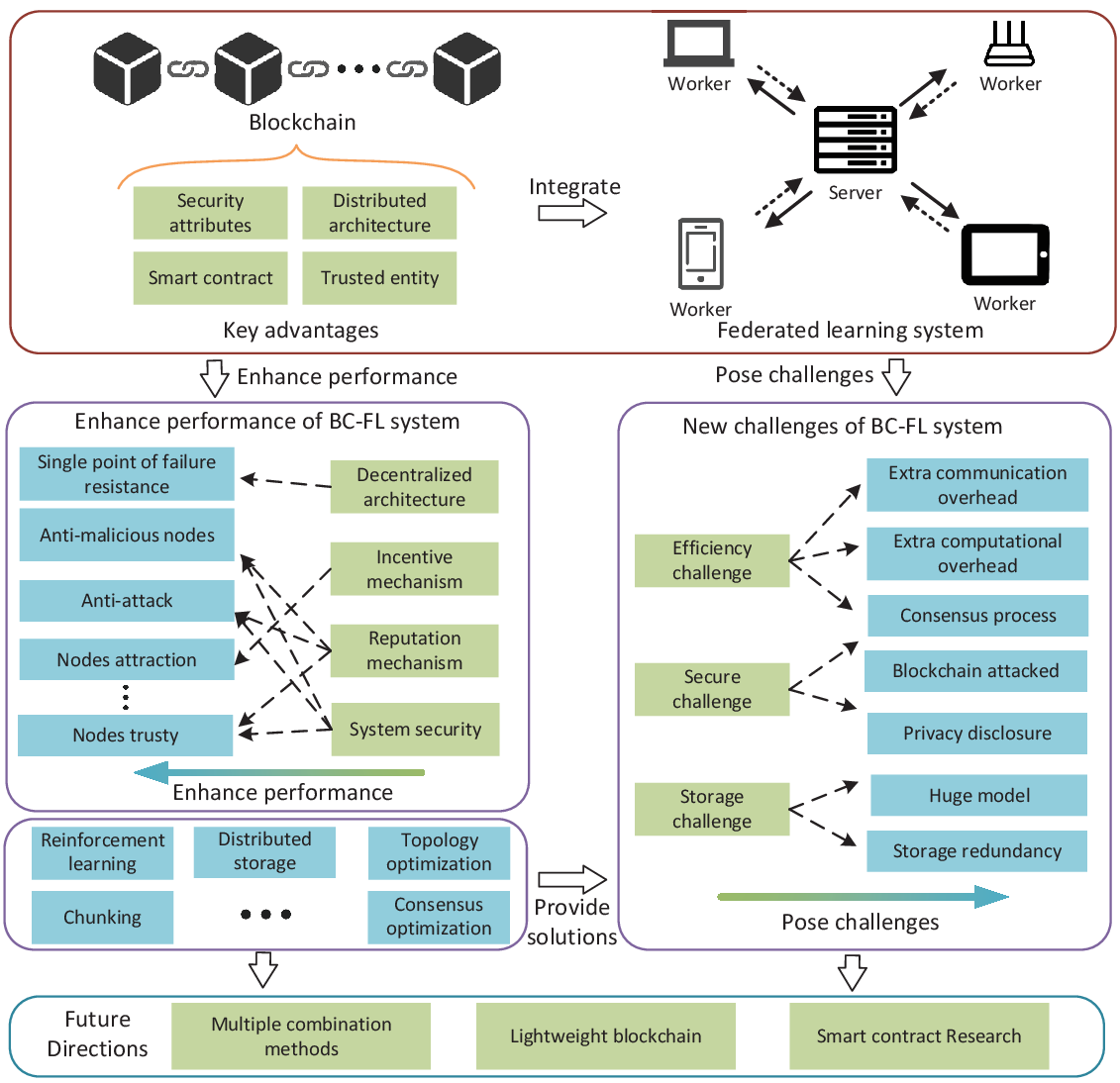}
    \caption{Main scope of this survey.
        We begin by exploring the characteristics of blockchain and investigate its enhancement of Federated Learning systems.
        Next, we discuss the additional challenges introduced by using blockchain in FL systems and review existing solutions.
        Finally, we outline future research directions for Blockchain-empowered Federated Learning systems.}
    \label{fig_total}
    \vspace{-15pt}
\end{figure*}

The introduction of blockchain has further driven the development of FL, but blockchain is not a panacea for FL.
Our research indicates that blockchain integration poses challenges related to runtime efficiency and storage capacity.
First, the consensus mechanism of blockchain adds communication and computation overhead to the BC-FL system.
Second, due to the distributed storage nature of blockchain, full nodes need to back up the entire blockchain data.
Additionally, the introduction of blockchain can also bring additional security issues, such as Sybil attacks \cite{hafid2022tractable}.

Currently, several surveys on BC-FL systems have been published.
Some focus on the integration of BC-FL with other fields, such as the Internet of Things, drones, and healthcare.
These studies emphasize the specific applications of BC-FL systems rather than their commonalities.
Other surveys investigate BC-FL systems in general.
Qu \emph{et al.} conducted a detailed study on the performance of decentralization, attack resistance, and incentive mechanisms in BC-FL systems, and surveyed the system architecture forms of BC-FL.
However, they did not investigate transparent reputation mechanisms in BC-FL and thoroughly analyze why blockchain can enhance FL systems, merely classifying the functions of BC-FL.
Zhu \emph{et al.} divided BC-FL system models into three categories and surveyed real-world applications of BC-FL.
However, they lacked a comprehensive investigation of single point of failure, reputation mechanisms, security, and privacy issues.
Additionally, the aforementioned surveys neglected to conduct a detailed investigation into the negative effects blockchain can bring to BC-FL systems.
Sameera \emph{et al.} summarized the general architecture of BC-FL and detailed how BC-FL addresses security and privacy threats.
However, they lacked in-depth research on BC-FL's reputation mechanisms, incentive mechanisms, and system efficiency and storage issues.
We believe that the various enhancements and potential challenges blockchain brings to FL systems stem from certain properties or functionalities of blockchain.
Meanwhile, some properties of blockchain can play different roles in FL systems depending on their application.
The contributions of this work are as follows:

\begin{itemize}
    \item Starting from the characteristics and functionalities of blockchain, we introduce how blockchain enhances FL systems in terms of decentralization, reputation mechanisms, incentive mechanisms, and security.
    \item We comprehensively investigate the additional challenges that arise from using blockchain in FL systems, the reasons behind these challenges, and existing solutions.
    \item We summarize future research directions for blockchain-based FL systems based on existing research.
\end{itemize}


Investigating how blockchain enhances FL systems from the perspective of blockchain's characteristics is an unaddressed area in existing surveys.
This survey can complement similar recent surveys, filling a gap in the research on BC-FL.
This survey.'s main scope is illustrated in Fig. \ref{fig_total}.
Section \ref{background} provides relevant background knowledge on FL, blockchain.
Section \ref{bc_fl_empower} details general BC-FL architecture and how blockchain enhances FL systems.
Section \ref{bc_fl_challenges} elaborates on the challenges of using blockchain in FL systems and existing solutions.
Section \ref{future_research_directions} we point out future research directions for BC-FL systems.
Section \ref{conclusion} concludes the paper.

\begin{figure*}
    \centering
    \includegraphics[width=5.0in]{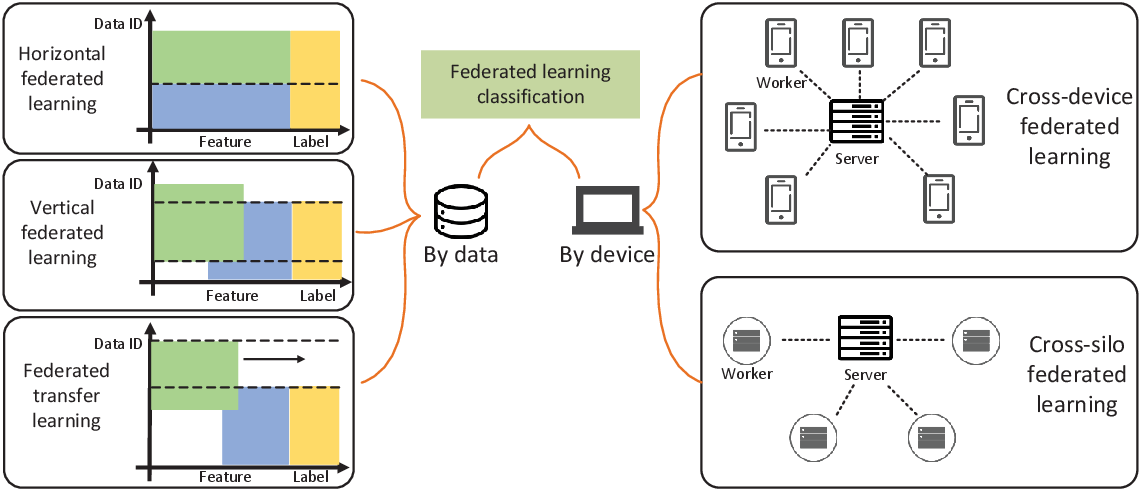}
    \caption{Categories of FL systems.}
    \label{fig_fl_class}
    \vspace{-15pt}
\end{figure*}

\section{Background}
\label{background}
\subsection{Federated Learning}
\begin{figure}
    \centering
    \includegraphics[width=4.0in]{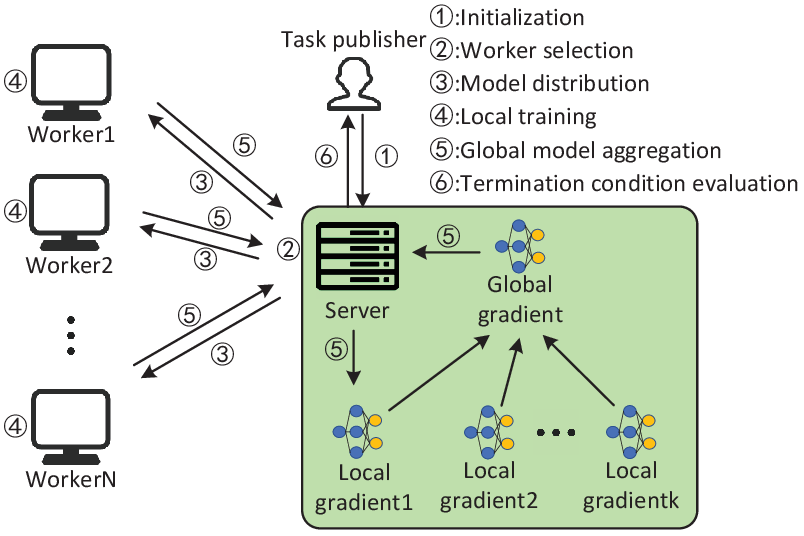}
    \caption{Standard federated learning (FL) training methodology.}
    \label{fig_fl_overview}
    \vspace{-15pt}
\end{figure}

FL is a privacy-preserving distributed machine learning paradigm proposed by Google \cite{mcmahan2017communication}.
This paradigm involves a network of multiple participants (clients) alongside a central server.
The clients are tasked with developing local models, which are then consolidated by the server into a unified, global model \cite{vargaftik2022eden,wang2022communication,zhang2021survey}.
This structure allows participants considerable autonomy, enabling them to contribute to the FL framework without disclosing their data to any FL node.
Participation in FL training remains at the discretion of the users.
Fig. \ref{fig_fl_overview} visually represents the standard FL training methodology.
The task initiator selects an FL server to publish the training task.
Subsequently, clients related to the training task join the FL training, and the server initializes the global model.
In each communication round, the server selects clients to participate in that round of training and distributes the global model to these clients.
The clients then use their local datasets to train the global model, resulting in local models, which they send back to the server.
The server aggregates these local models into a new global model according to certain rules.
The FL process stops when the training termination condition is met; otherwise, training continues.

FL generally has two classification methods, as depicted in Fig. \ref{fig_fl_class}.
Based on the sample ID and feature distribution of local datasets, FL systems can be divided into Horizontal FL (HFL), Vertical FL (VFL), and Federated Transfer Learning (FTL).
In HFL, users have different sample IDs but identical data features.
In VFL, users have identical sample IDs but different data features.
In FTL, both the sample IDs and data features differ.
Generally, current research on BC-FL is predominantly based on HFL, with only a few studies on VFL \cite{teimoori2022secure, xu2021method} and FTL \cite{zhang2021federated}.
Despite the datasets' different characteristics, blockchain's role does not fundamentally differ.

Moreover, based on the different types of user devices involved in training, FL can be divided into cross-device FL and cross-silo FL.
Cross-device FL involves an extensive array of mobile and IoT devices, potentially numbering up to $10^{10}$, often with poor device performance and network conditions \cite{chen2023fs,dorfman2023docofl}.
In cross-silo FL systems, the participating clients are typically professional computing nodes maintained by specialized institutions, and the number of clients is generally fewer than 100 \cite{huang2023promoting,yuan2023adaptive}.

\subsection{Blockchain}
\begin{figure}
    \centering
    \includegraphics[width=4.0in]{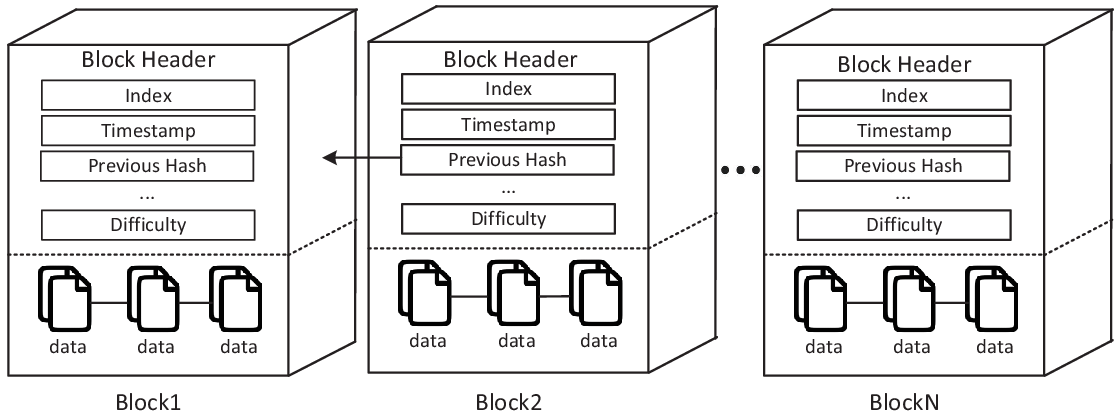}
    \caption{The structure of a blockchain.}
    \label{fig_blockchain}
    \vspace{-15pt}
\end{figure}

Blockchain is a distributed ledger system designed to securely, transparently, and immutably record data in a decentralized manner \cite{zheng2018blockchain,ye2023survey}.
It is resilient to the presence of some malicious nodes and is resistant to control by any single entity \cite{liu2023flexible,chen2023vehicular}.
A blockchain is essentially composed of interconnected blocks arranged sequentially in a chronological chain.
This structure is illustrated in the Fig. \ref{fig_blockchain}.
Each block includes a header and a body.
The block header typically contains metadata such as the block number, timestamp, and the hash of the previous blocks.
The specific parameters stored in the header can vary depending on the blockchain system.
The block body stores specific information, like transactions.

Blockchain can be classified into two main types based on node participation restrictions: permissionless blockchain and permissioned blockchain.
The permissionless blockchain allows unrestricted node participation in the system's operations, exemplified by Bitcoin and Ethereum.
These blockchains operate without the need for approvals or authorizations from central authorities.
In contrast, the permissioned blockchain is managed by a specific organization, and only authorized nodes can access the system.
It is suitable for data privacy and security applications, such as financial and government information management.

The autonomous operation of blockchains in the absence of central oversight relies on consensus mechanisms.
These algorithms define the state of the blockchain and are vital for its functionality.
Broadly, consensus mechanisms are categorized into proof-based and committee-based systems \cite{xu2023survey}.
Proof-based systems prioritize nodes based on specific resource possession; for instance, Proof of Work (PoW) and Proof of Stake (PoS) are notable examples \cite{lepore2020survey}.
In PoW, nodes compete to solve complex puzzles, with computational prowess conferring a higher accounting priority.
PoS, on the other hand, employs an encrypted random selection algorithm to appoint a leader for block creation, with a node's selection likelihood tied to its token holdings.
Proof-based consensus mechanisms are predominantly utilized in permissionless blockchains due to their robust defence against malicious nodes.
Conversely, committee-based mechanisms utilize a voting process where consensus on a new block's addition is reached through a predetermined number of affirmative votes.
Protocols such as Raft \cite{huang2019performance} and PBFT \cite{li2020scalable} exemplify committee-based mechanisms.
Committee-based mechanisms generally offer higher throughput than proof-based mechanisms and are often used in permissioned blockchains.
However, they can be communication-intensive, posing challenges in large-scale networks.

\section{Blockchain-empowered Federated Learning}
\label{bc_fl_empower}
This section introduces how blockchain technology enhances the FL system.
These enhancements can be categorized into four aspects: decentralization, reputation evaluation mechanism, incentive evaluation mechanism, and security.
Each of these four aspects will be discussed separately.

\subsection{Decentralization}
FL traditionally relies on a central parameter server, where clients must continuously communicate with a single FL server.
This centralized structure poses significant risks, such as single points of failure and potential malicious server behaviour.
Furthermore, node reputation information is solely managed by the server, which is not ideal for developing an open FL ecosystem.
Blockchain's decentralization is a core feature that fundamentally addresses these issues and provides a new architectural approach to FL systems.
Decentralization can enhance the security, transparency, and reliability of FL systems, with these benefits manifesting in various facets of blockchain's advantages for FL.

Our review of existing literature identifies several factors influencing the decentralization of BC-FL, including system architecture, consensus mechanisms, and smart contracts.
The system architecture determines which nodes maintain the blockchain, while the consensus mechanism dictates which nodes have the right to manage the system.
Smart contracts can automate various algorithms within the BC-FL system, offering greater scalability.
Table \ref{table:decentralization} compiles representative BC-FL systems, detailing their use of smart contracts, architecture, consensus mechanisms, and experimental platforms.

\begin{table}
  \setlength{\abovecaptionskip}{0pt}
  \setlength{\belowcaptionskip}{0pt}
  \caption{BC-FL Systems Based on Blockchain and Federated Learning.}
  \label{table:decentralization}
  \scriptsize
  \renewcommand{\arraystretch}{1.2}
  \centering
  \tabcolsep 2pt
  \resizebox{0.8 \linewidth}{!}{
  \begin{tabular}{lcccc}
      \toprule
      Ref.                              & Smart contract & Architecture & Consensus algorithm & Platform           \\
      \midrule
      Abdel \cite{abdel2022privacy}     & $\checkmark$   & Complete     & Algorand            & Other              \\
      Fang \cite{fang2022privacy}       & $\times$       & Partial      & Algorand            & Other              \\
      Feng \cite{feng2021blockchain}    & $\checkmark$   & Partial      & PBFT/Raft           & Hyperledger Fabric \\
      Guo \cite{guo2022sandbox}         & $\checkmark$   & Partial      & PBFT                & Hyperledger Fabric \\
      Jiang \cite{jiang2021cooperative} & $\times$       & Complete     & DPoS                & Other              \\
      Liu \cite{liu2021blockchain}      & $\times$       & Partial      & PoW+PoA             & Other              \\
      Lu \cite{lu2020blockchain}        & $\times$       & Complete     & DPoS                & Other              \\
      Nguyen \cite{nguyen2021federated} & $\times$       & Partial      & PoR                 & Other              \\
      Nguyen \cite{nguyen2022latency}   & $\times$       & Partial      & PoW                 & Other              \\
      Qi \cite{qi2022high}              & $\checkmark$   & Partial      & -                   & Hyperledger Fabric \\
      Qi \cite{qi2021blockchain}        & $\checkmark$   & Complete     & Modified PBFT       & Ethereum           \\
      Qu \cite{qu2020blockchained}      & $\times$       & Complete     & PoW                 & Other              \\
      Rehman \cite{Rehman2021trustfed}  & $\checkmark$   & Complete     & -                   & Ethereum           \\

      Wu \cite{wu2022a}                 & $\times$       & Complete     & PoW                 & Other              \\
      Xu \cite{xu2021bafl}              & $\times$       & Complete     & -                   & Other              \\
      Xu \cite{xu2021besifl}            & $\checkmark$   & Complete     & -                   & Other              \\
      Zhang \cite{zhang2021bc}          & $\times$       & Partial      & PoW                 & Other              \\
      Zhao \cite{zhao2021deep}          & $\times$       & Partial      & PBFT                & Other              \\
      Wang \cite{wang2023incentive}     & $\times$       & Complete      & -                  & Other \\
      Huang \cite{wang2023incentive}     & $\checkmark$    & Partial      & Raft, HotStuff  & FISCO \\
      Ouyang \cite{ouyang2023artificial}     & $\checkmark$    & Partial      & PoS  & Ethereum \\
      Yuan \cite{yuan2024secure}     & $\checkmark$    & Partial      & Raft, Modified DAG  & Hyperledger, DAG \\
      aloqaily \cite{aloqaily2023reinforcing}     & $\times$    & Partial      & -  & Other \\
      Mu \cite{mu2023digital}     & $\times$    & Complete      & -  & Other \\
      Wahrstatter \cite{wahrstatter2024openfl}     & $\times$    & Complete      & PoS  & Ethereum \\
      \bottomrule
  \end{tabular}
  }
  \vspace{-15pt}
\end{table}

\begin{figure*}
    \centering  
    \subfigure[Complete decentralization.]{
        \label{fig_complete}
        \includegraphics[width=2.5in]{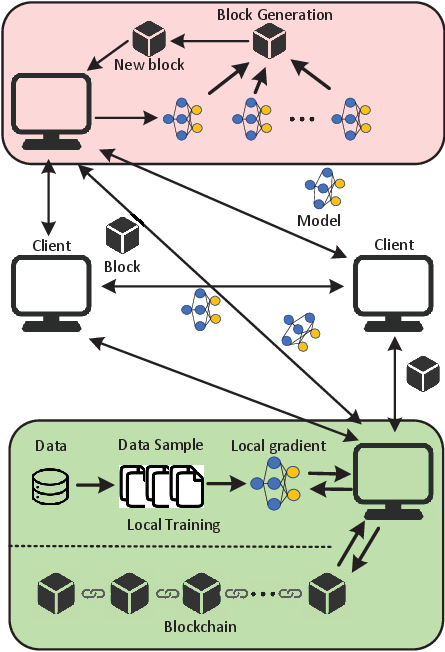}}
    \subfigure[Partial decentralization.]{
        \label{fig_incomplete}
        \includegraphics[width=2.5in]{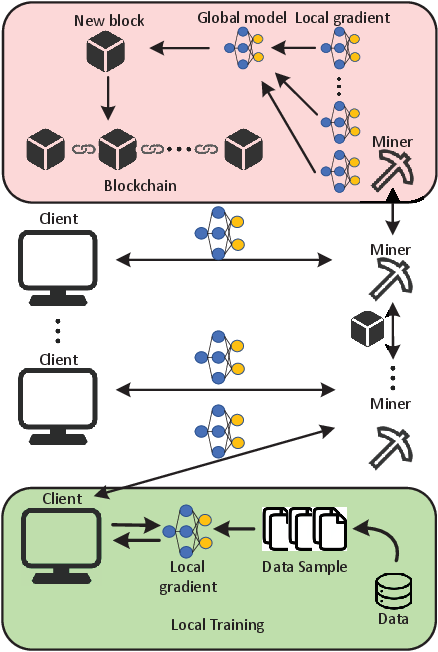}}
    \vspace{-.1in}
    \caption{Two decentralized architectures of the Blockchain-Empowered Federated Learning system.}
    \label{fig009}
    \vspace{-15pt}
\end{figure*}

\textbf{Architecture.}
BC-FL systems can be categorized by their degree of decentralization: complete and partial.
In the completely decentralized BC-FL, all nodes are eligible to participate in the consensus process of the blockchain.
Fig. \ref{fig_complete} shows its general system architecture.
This approach demands high computational and storage capacities from all nodes.
Conversely, partially decentralized BC-FL involves only a subset of nodes running the blockchain, while others focus solely on FL training.
Fig. \ref{fig_incomplete} shows its general system architecture.
The selected nodes that operate the blockchain system are known as super nodes and typically have stronger computing power and better communication conditions.
This approach sacrifices some transparency for increased efficiency.

\textbf{Consensus Mechanism.}
A considerable portion of the work adopts common blockchain consensus mechanisms such as PoW and PBFT.
PoW involves blockchain nodes competing to solve a mathematical problem, with the first solver aggregating models and training information into a new block.
Other nodes then verify the block's correctness, and upon majority approval, it is added to the blockchain.
In PBFT, a set of consensus nodes is chosen within the BC-FL system, from which a leader node aggregates the model and generates a new block.
Other nodes in the set verify the leader's block.
Some work has specifically developed consensus mechanisms for BC-FL, such as Proof of Reputation \cite{nguyen2021federated}. These custom consensus mechanisms are usually designed to enhance FL functionality or mitigate the disadvantages of blockchain, which will be elaborated on in the subsequent sections.

\textbf{Smart Contracts.}
Smart contracts significantly enhance the scalability of BC-FL systems.
For instance, model aggregation can be executed via smart contracts, increasing transparency.
Additionally, smart contracts can deploy algorithms for detecting and handling malicious nodes, thereby improving system efficiency.
They can also manage node reputation evaluations and incentive algorithms, further enhancing system transparency.

Next, we will examine some representative architectures of completely decentralized BC-FL systems.

In \cite{xu2021besifl}, Xu \emph{et al.} proposed a BC-FL framework named Blockchain Empowered Secure and Incentive Federated Learning (BESIFL).
BESIFL enables any node in the network to initiate FL training requirements.
Upon receipt of a requirement, BESIFL selects computing nodes with high computation reputation scores to form a computing pool and assigns them the task of model training.
Meanwhile, BESIFL chooses verification nodes with high verification reputation scores to form a verification pool and assigns them the task of model aggregation and verification using pre-defined procedures specified by the smart contract.
Li \emph{et al.} also proposed a completely decentralized BF-FL system, where each client acts as both a FL trainer and a blockchain miner \cite{li2021blockchain}.
After training their local models, clients initiate blockchain transaction requests and broadcast their models by attaching them to the transaction information.
Each client aggregates the global model locally after receiving local models from all other clients and starts mining.
The winning miner broadcasts a block containing global model information, which is verified by other clients and then written into the blockchain.
However, this system assumes that all clients possess equal computational power, which may not be realistic in practice.
In addition to the two aforementioned decentralized methods, Qu \emph{et al.} designed a novel approach that utilizes a rotation mechanism with randomness to select committee members for participating in blockchain consensus \cite{qu2022fl}.
This proposed blockchain consensus mechanism greatly reduces additional consumption generated by the blockchain consensus process compared to the PoW mechanism.
Committee members are only responsible for aggregating and validating the global model and do not participate in training.
The global model is generated by committee members and stored in the blockchain after verification.
While the rotation mechanism ensures the mobility of committee members, it can ensure some level of system security.
However, this consensus mechanism is only applicable in situations where the number of malicious nodes is small.

The BC-FL systems described below follow the partial decentralization architecture.
Feng \emph{et al.} proposed a BC-FL system for UAVs that maintains the blockchain system only in entities with high computing and storage capabilities, such as base stations and roadside nodes \cite{feng2021blockchain}.
This approach enables transparent and automated model aggregation operations through the use of smart contracts, which replace the traditional parameter server.
To address the challenge of online and offline state changes among BC-FL participants, the authors set the maximum waiting time and the required number of local models for each learning round.
If any of these conditions are met, the model update contract is triggered, ensuring timely updates while accommodating BC-FL participant availability.
In  \cite{liu2021blockchain}, Liu \emph{et al.} proposed a framework for training vehicle intrusion model.
The blockchain is maintained by roadside units and stores and shares the global models for the BC-FL system.
After receiving the global model, the vehicle uses the data collected by itself to train the model and upload it to the connected roadside unit nodes.
The consensus mechanism in place combines PoW and PoA, with the roadside node that has achieved the highest accuracy being written into the block to encourage the training of high-precision models.

\textbf{Workflow of BC-FL Systems.}
The overall workflow of the BC-FL system is illustrated in Fig. \ref{fig_BC-FL}.
Different BC-FL systems may be adjusted according to specific circumstances.
The steps are explained as follows:

\begin{figure*}
    \centering
    \includegraphics[width=4.5in]{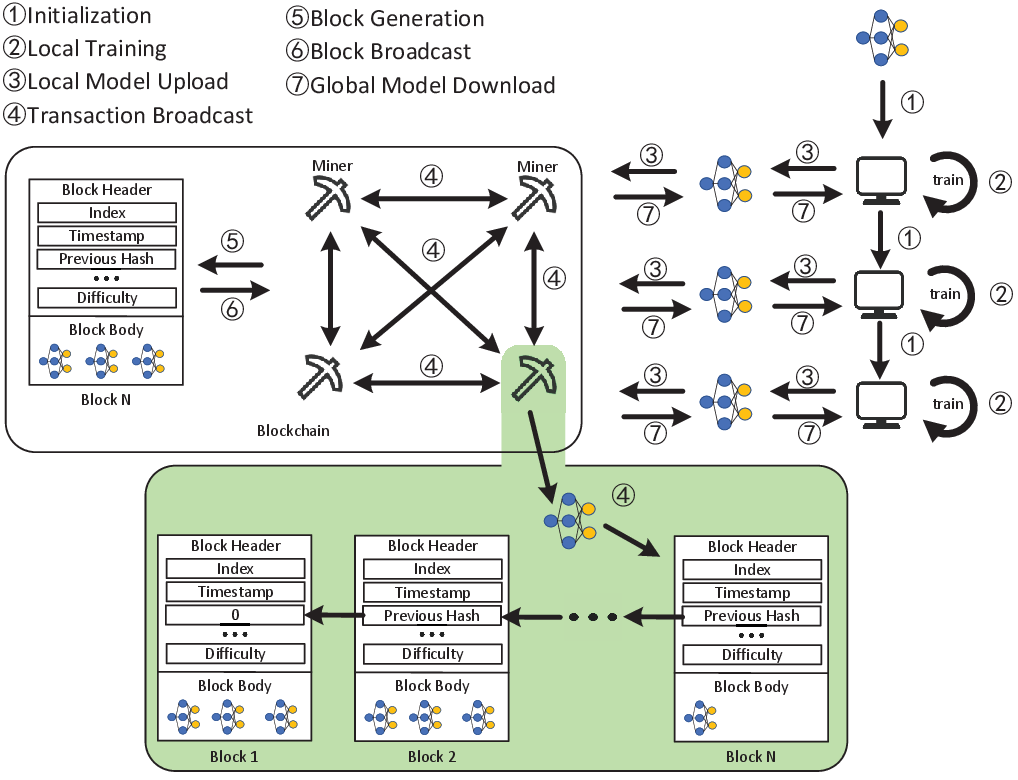}
    \vspace{-.1in}
    \caption{Overall structure and workflow of the blockchain-empowered federated learning system.}
    \label{fig_BC-FL}
    \vspace{-15pt}
\end{figure*}

Step 1. Initialization:
Each client initializes the environment based on prior negotiation, including model parameter, and training parameter.
The blockchain can assist clients in negotiation by storing initialization parameters on the chain and using smart contracts.

Step 2: Local model training.
Each client trains the global model using their local dataset.

Step 3: Local model upload.
Clients upload training-related data and local models to the blockchain system.
To alleviate storage pressure on the blockchain, clients may upload only model-related information rather than the entire model, as detailed in Section \ref{storage_challenges_and_solutions}.

Step 4: Transaction broadcast.
Upon receiving the transaction, blockchain nodes broadcast it within the system for cross-validation.
The nodes inspect the transaction content (e.g., model) based on pre-defined rules.
If no issues are found, the transaction is added to the local transaction pool.

Step 5: Block generation.
Blockchain nodes select the node with the right to generate blocks for the current round based on the consensus protocol.
This node aggregates the local models to generate the global model, compiles relevant model and training information, and creates a block.

Step 6: Block Broadcast.
The blockchain system broadcasts the newly generated block.
Upon receiving it, validation nodes verify the block according to specific rules.
If the majority of nodes validate the block, it is added to their locally maintained blockchain, achieving consensus across the network.

Step 7: Global Model Download.
Clients download the latest global model from the blockchain system.

Step 8: End condition judgment.
Based on pre-negotiated rules, the FL process evaluates whether it has reached the end condition.
If not, the process returns to Step 2 to continue training.

\subsection{Reputation Evalutation Mechanism}

FL is a collaborative approach to training a shared model that requires the participation of multiple clients with local data.
However, clients may have varying motivations and behaviors, such as seeking rewards for their assistance, hoping to obtain a trained model, or attempting to benefit from the global model without contributing to the training process.
In some cases, clients may even have malicious intentions, seeking to undermine the effectiveness of FL due to conflicts of interest in reality or other factors.
Compared to traditional distributed learning methods, FL prioritizes user data privacy, which means that the parameter server has limited access to information about the local environment of each client.
Therefore, it is essential for the FL task publisher to implement a reputation management mechanism that can assist in managing, rewarding, or punishing FL clients based on their contributions and behavior.

Several studies have proposed the use of some reputation management mechanisms in a centralized way on the parameter server \cite{he2022cgan, guo2022sandbox}.
While this approach can serve as a foundation for client management, reward and punishment schemes, its lack of transparency remains a concern.
Data owners who contribute to the training process may worry about potential inaccuracies in the parameter server's reputation calculations, while those seeking to obtain a trained model may be concerned that the parameter server could intentionally manipulate reputations to undermine FL models.
Given the importance of attracting high-quality data owners to ensure optimal FL model performance, the transparent reputation management mechanism is particularly well-suited for FL systems.
Additionally, a trustworthy parameter server aims to calculate reputation in a transparent manner to discourage malicious nodes.
To address these concerns, the BC-FL system leverages blockchain technology to ensure the transparency and credibility of the reputation management mechanism.

After conducting our analysis, we have identified two crucial functions that blockchain can perform within the reputation management mechanism.
\begin{itemize}
    \item [1.]The blockchain acts as a reliable third-party ledger in the BC-FL system to document crucial information regarding each node's reputation, including but not limited to its reputation value \cite{guo2022sandbox,kang2019incentive,haddaji2022federated} and various calculation bases \cite{kang2021optimizing,liu2021blockchain,qi2022high}.
    \item [2.]In the BC-FL system, the reputation computation process can be deployed on the blockchain through a specialized reputation calculation smart contract \cite{kang2021optimizing,qiu2022rendering,qi2022high}.
          This approach serves to ensure both transparency and automation throughout the entire computation process, thereby guaranteeing dependable and consistent outcomes.
\end{itemize}

\begin{figure}
  \setlength{\abovecaptionskip}{0pt}
  \setlength{\belowcaptionskip}{0pt}
  \centering
  \includegraphics[width=4.5in]{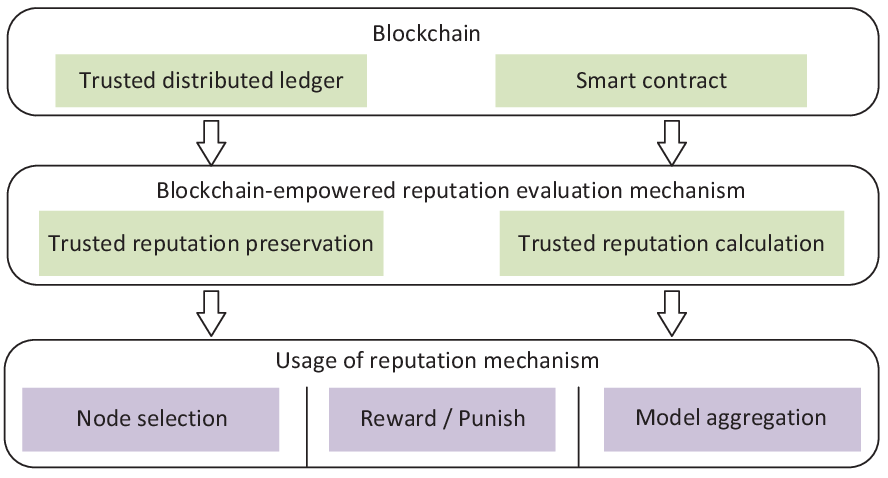}
  \caption{Reputation management mechanisms based on blockchain.
      Blockchain is commonly utilized as a reliable distributed ledger or transparent smart contract platform for reputation management mechanisms.
      This allows the system to store clients' reputation value and the reputation calculation basis on the blockchain, or use smart contracts to compute the reputation in a transparent way.
      The primary function of reputation management mechanisms is to facilitate node selection, model aggregation, and incentivization.}
  \label{fig_reputation}
  \vspace{-15pt}
\end{figure}

The reputation management mechanism based on blockchain in BC-FL is illustrated in Fig. \ref{fig_reputation}.
Our study of recent research papers on client reputation calculation methods has revealed that the multiweight subjective logic calculation method is a popular choice for enhancing the trustworthiness and reliability of BC-FL systems.
To elucidate the operation of the reputation management mechanism in BC-FL systems, we will present a concise overview of the multi-weight subjective logic calculation method.

This method aims to assess the reputation value of a client by considering three crucial attributes: positive evaluation, negative evaluation, and uncertain evaluation.
For example, in  \cite{kang2019incentive,kang2020reliable,kang2021optimizing}, Kang \emph{et al.} demonstrated how multi-weight subjective logic can be used to accurately calculate reputation values.
In their proposed BC-FL systems, the task publisher, denoted as $TP_i$, calculates the reputation of each client through two main components.
The first part involves direct reputation calculation, where $TP_i$ evaluates BC-FL clients based on three attributes: belief, disbelief, and uncertainty, corresponding to positive evaluation, negative evaluation, and uncertain evaluation, respectively.
To facilitate comprehension, we simplify the formula as follows:
\begin{equation}
    \begin{cases}
        b^{dir}_{i \rightarrow j} =  (1-u_i)\frac{a_i}{a_i+b_i} \\
        d^{dir}_{i \rightarrow j} =  (1-u_i)\frac{b_i}{a_i+b_i} \\
        u^{dir}_{i \rightarrow j} = 1-q_{i \rightarrow j}
    \end{cases},
\end{equation}
where $b^{dir}_{i \rightarrow j}$, $d^{dir}_{i \rightarrow j}$, and $u^{dir}_{i \rightarrow j}$ need to satisfy the restrictions:
\begin{equation}
    \begin{split}
        b^{dir}_{i \rightarrow j}+d^{dir}_{i \rightarrow j}+u^{dir}_{i \rightarrow j} &=1,\\
        0 \le b^{dir}_{i \rightarrow j} \le 1, ~~0 \le d^{dir}_{i \rightarrow j} \le 1,&  ~~0  \le u^{dir}_{i \rightarrow j} \le 1.
    \end{split}
\end{equation}

The variables $a_i$ and $b_i$ represent the positive and negative evaluations of client $C_j$ by $TP_i$ respectively.
Variables $b^{dir}_{i \rightarrow j}$, $d^{dir}_{i \rightarrow j}$, and $u^{dir}_{i \rightarrow j}$ correspond to the previously mentioned belief, disbelief, and uncertainty.
Variable $q_{i \rightarrow j}$ denotes the probability of successful delivery of data packets sent by $C_j$ to $TP_i$.
The direct reputation $DIR_{i \rightarrow j}$ is then expressed as:
\begin{equation}
    DIR_{i \rightarrow j}=b^{dir}_{i \rightarrow j}+\alpha u^{dir}_{i \rightarrow j},
\end{equation}
where $\alpha$ is a variable factor between 0 and 1.

The second part involves the evaluation of the client's reputations by other task publishers $TP_k$.
First, $TP_k$ sends its reputation opinion vector to $TP_x$, which then calculates the credibility of $TP_k$'s reputation opinion through an amendatory cosine function.
Then, form the weight of publisher $k$ by calculated credibility.
All task publishers' reputation opinions are combined in a weighted manner to form an indirect reputation.

Combining direct and indirect reputations results in the final value of belief $b^{final}_{i \rightarrow j}$, disbelief $d^{final}_{i \rightarrow j}$ and uncertainty $u^{final}_{i \rightarrow j}$ of the $C_j$, as defined as:
\begin{equation}
    \begin{cases}
        b^{final}_{i \rightarrow j} = \frac{b^{dir}_{i \rightarrow j}u^{rec}+b^{rec}u^{dir}_{i \rightarrow j} }{u^{dir}_{i \rightarrow j}+u^{rec}-u^{rec}u^{dir}_{i \rightarrow j}} \\
        d^{final}_{i \rightarrow j} = \frac{d^{dir}_{i \rightarrow j}u^{rec}+d^{rec}u^{dir}_{i \rightarrow j} }{u^{dir}_{i \rightarrow j}+u^{rec}-u^{rec}u^{dir}_{i \rightarrow j}} \\
        u^{final}_{i \rightarrow j}=\frac{u^{rec}u^{dir}_{i \rightarrow j}}{u^{dir}_{i \rightarrow j}+u^{rec}-u^{rec}u^{dir}_{i \rightarrow j}}
    \end{cases},
\end{equation}
where $b^{rec}$, $d^{rec}$, and $u^{rec}$ are the belief, disbelief and uncertainty of the indirect reputation mentioned above.
Then, we compute the final reputation $REP_{i \rightarrow j}$ of $C_j$ in:
\begin{equation}
    REP_{i \rightarrow j}=b^{final}_{i \rightarrow j}+\alpha u^{final}_{i \rightarrow j}.
\end{equation}

\begin{table*}
    \setlength{\abovecaptionskip}{0pt}
    \setlength{\belowcaptionskip}{0pt}
    \caption{Blockchain-based reputation mechanism in BC-FL Systems.}
    \label{table:reputation}
    \scriptsize
    \renewcommand{\arraystretch}{1.2}
    \centering
    \tabcolsep 2pt
    \begin{tabular}{L{0.8in} C{0.6in} C{0.45in} C{0.55in} C{0.4in} C{0.4in} C{0.6in} C{0.5in} C{0.6in}  }
        \toprule
        \multirow{2}*{Ref.}                & \multicolumn{3}{c}{Reputation Source} & \multicolumn{2}{c}{Blockchain Usage} & \multicolumn{3}{c}{Reputation Usage}                                                                                           \\
        \cmidrule (lr){2-4}\cmidrule (lr){5-6}\cmidrule (lr){7-9}
                                           & Aggregration                          & Other Workers                        & Blockchain                           & Usage 1      & Usage 2      & Model Aggregation & Node Selection & Reward or Punishment \\
        \midrule
        Chen \cite{chen2022repbfl}         & $\checkmark$                          & $\checkmark$                         & -                                    & -            & -            & -                 & $\checkmark$   & -                    \\
        Gao \cite{gao2022fgfl}             & $\checkmark$                          & -                                    & -                                    & $\checkmark$ & $\checkmark$ & -                 & $\checkmark$   & $\checkmark$         \\
        Guo \cite{guo2022sandbox}          & $\checkmark$                          & -                                    & -                                    & -            & -            & -                 & $\checkmark$   & -                    \\
        Haddaji\cite{haddaji2022federated} & $\checkmark$                          & -                                    & $\checkmark$                         & $\checkmark$ & -            & -                 & $\checkmark$   & -                    \\
        He \cite{he2022cgan}               & $\checkmark$                          & -                                    & -                                    & -            & -            & -                 & $\checkmark$   & -                    \\
        Kang \cite{kang2021optimizing}     & $\checkmark$                          & -                                    & -                                    & $\checkmark$ & -            & -                 & $\checkmark$   & -                    \\
        Liu \cite{liu2021blockchain}       & $\checkmark$                          & -                                    & -                                    & $\checkmark$ & -            & -                 & $\checkmark$   & $\checkmark$         \\
        Qi \cite{qi2022high}               & $\checkmark$                          & $\checkmark$                         & -                                    & $\checkmark$ & $\checkmark$ & $\checkmark$      & -              & $\checkmark$         \\
        Qiu \cite{qiu2022rendering}        & $\checkmark$                          & -                                    & $\checkmark$                         & $\checkmark$ & $\checkmark$ & $\checkmark$      & -              & -                    \\
        Rahman \cite{rahman2020secure}     & $\checkmark$                          & -                                    & -                                    & $\checkmark$ & -            & -                 & $\checkmark$   & -                    \\
        Xu \cite{xu2021bafl}               & $\checkmark$                          & -                                    & -                                    & $\checkmark$ & $\checkmark$ & $\checkmark$      & $\checkmark$   & $\checkmark$         \\
        Zhao \cite{zhao2020privacy}        & $\checkmark$                          & -                                    & -                                    & $\checkmark$ & -            & -                 & $\checkmark$   & $\checkmark$         \\
        Wang \cite{wang2023incentive}     & $\checkmark$                          & -                                    & $\checkmark$                           & $\checkmark$ & -            & $\checkmark$                 & -   & -        \\
        Lin \cite{lin2023drl}     & $\checkmark$                          & -                                    & $\checkmark$                           & $\checkmark$ & -            & -                & $\checkmark$   & -        \\
        Fu \cite{fu2023incentive}     & $\checkmark$                          & -                                    & -                           & $\checkmark$ & -            & -                & -   & $\checkmark$        \\
        Wahrstatter \cite{wahrstatter2024openfl}     & $\checkmark$                          & -                                    & -                           & $\checkmark$ & $\checkmark$            & -                & -   & $\checkmark$        \\
        \bottomrule
    \end{tabular}
    \vspace{-15pt}
\end{table*}

Various papers adopt distinct approaches in calculating the reputation of BC-FL clients.
Some calculate reputation values solely on the basis of local model test accuracy, while others take into account evaluations from other clients or factor in the interaction effect between clients and the blockchain system.
Moreover, researchers have leveraged clients' reputations in various ways.
For instance, some deploy reputation as a criterion for selecting participating clients, whereas others utilize it to ascertain the weight assigned to global model aggregation.
Additionally, there are those who offer incentives and penalties to clients based on their respective reputations.

We present a comprehensive analysis of BC-FL systems that utilize blockchain technology to establish transparent reputation management mechanisms.
Table \ref{table:reputation} summarizes the key attributes of these systems.

The attributes ``Aggregation'', ``Other Workers'', and ``Blockchain'' represent the basis for evaluating the reputation of clients.
A checkmark in the corresponding box signifies that the BC-FL system takes the attribute into consideration when calculating the client's reputation.
``aggregation'' represents the contribution of a client's local model towards the global model during aggregation, including evaluating the accuracy of the local model.
``other workers'' relates to the interaction between clients, specifically through peer evaluation among training clients.
``Blockchain'' encompasses the effect of a client's participation in blockchain maintenance activities, such as successful block generation.
The attributes ``Usage 1'' and ``Usage 2'' illustrate two potential roles that blockchain may play in the node reputation mechanism, as previously mentioned.
``Usage 1'' describes blockchain's involvement in the BC-FL system's reputation management mechanism as a transparent and open ledger.
``Usage 2'' involves the use of smart contracts to automatically and transparently calculate a client's reputation.
In addition, we examine how reputation is utilized in BC-FL systems by exploring the attributes of ``Model aggregation'', ``Node Selection'', and ``Reward or Punishment``.
``Model aggregation'' involves weighting a local model based on the client's reputation value when aggregating the global model.
``Node selection'' indicates that the FL client selection process in each round will consider its reputation value.
``Reward or punishment'' signifies the use of reputation value as the basis for rewarding or punishing the client.

In addition to the reputation calculation methods discussed earlier, several other approaches have been proposed in the literature.
In  \cite{qi2022high}, Qi \emph{et al.} proposed a novel reputation evaluation mechanism for multi-model aggregators in FL.
Each model aggregator has its test dataset, and the reputation of each participating client is calculated separately by each aggregator.
The winning aggregator is selected based on a set of rules, and the winning aggregator updates the client's reputation value to the blockchain.
The model aggregators calculate the client's reputation in two steps.
In the first step, each model aggregator uses a fair-value game \cite{gollapudi2017profit} to test the quality of the local model with its test dataset.
When the result of a formula containing model test accuracy reaches a certain threshold, the corresponding reputation update is activated.
In the second step, the model aggregator synthesizes the results given by other aggregators on the network to obtain the indirect reputation value of the node.
Finally, the reputation evaluation value of the modified model aggregator for the node in this round is obtained from the results of the first and second steps.
This approach ensures fairness in reputation evaluation across different aggregators and improves the accuracy of the final reputation value.

In  \cite{gao2022fgfl}, Gao \emph{et al.} designed a time-decaying subjective logic model  (SLM) algorithm to measure the client's reputation and a lightweight approach based on gradient similarity to measure client contribution.
The final task publisher determines the client's reward share by multiplying the contribution and reputation metrics.
They used reputation metrics to measure client reliability and select clients with high reputations to ensure high system stability, which enables their proposed system to work stably in unreliable environments.

\subsection{Incentive Evaluation Mechanism}
In FL systems, clients not only need to contribute local data but also consume significant amounts of computing resources and network bandwidth \cite{yu2020fairness,zhan2021survey}.
Without tangible incentives, it may be difficult to attract enough clients to participate in the FL systems.
Therefore, introducing an incentive mechanism in FL systems is critical.
The introduction of incentives can help incentivize clients to join the FL systems and contribute their valuable data.
Adequate participation is crucial for FL to train accurate models with good generalization.
Additionally, incorporating incentives can increase clients' engagement and motivation, leading to contributing better data and participation in more training epochs \cite{zhan2020learning}.
Furthermore, the incentive mechanism can help achieve fairness in FL systems by rewarding clients based on their data quality and computing power.

A transparent and open incentive mechanism is crucial for attracting clients to participate in federated learning.
As it involves vital interests, each client hopes to supervise the calculation of rewards.
The BC-FL system utilizes the blockchain to provide a transparent and open incentive mechanism.
The blockchain is a decentralized ledger that is maintained on each participating node, requiring the joint efforts of blockchain nodes instead of a centralized organization.
This architecture ensures transparency and openness and facilitates tracking and auditing of data necessary for calculating incentives, thereby establishing clients' trust in the incentive results.
Furthermore, the incentive algorithm can be written as a smart contract and deployed on the blockchain for automatic incentive calculation and distribution, further strengthening clients' trust in the incentive results.
The transparent and open incentive mechanism provided by the blockchain can help to attract more clients to participate in the FL process, contributing high-quality data and computing resources.
Consequently, it promotes the accuracy and generalization of the trained model and enhances the efficiency of the BC-FL system.

We focus on BC-FL systems that provide transparent and open incentive mechanisms based on the blockchain.
We believe that understanding this incentive mechanism requires consideration of three aspects: incentive basis, incentives, and incentive algorithms.
The settings of these aspects should be tailored to the specific FL tasks.
Table \ref{table:incentive} outlines several prominent BC-FL systems developed in recent years.

\begin{table*}
    \caption{Statistics of Blockchain Based Incentive Mechanism in BC-FL Systems.}
    \label{table:incentive}
    \scriptsize
    \renewcommand{\arraystretch}{1.2}
    \centering
    \tabcolsep2pt
    \begin{tabular}{L{0.8in}  L{1.2in}  L{2.5in} C{0.8in}}
        \toprule
        Ref.                             & Incentives              & Incentive Basis                                                 & Smart Contract \\
        \midrule
        Abdel \cite{abdel2022privacy}    & Manufacturer's discount & Reputation                                                      & $\checkmark$   \\
        Chen \cite{chen2022dimds}        & Token                   & Model accuracy                                                  & $\checkmark$   \\

        Gao \cite{gao2022fgfl}           & Token                   & Model accuracy, reputation                                      & $\checkmark$   \\
        Li \cite{li2021byzantine}        & Token                   & Model accuracy                                                  & -              \\
        Liu \cite{liu2021blockchain}     & Reputation              & Model accuracy                                                  & -              \\
        Liu \cite{liu2020secure_5g}      & Ethereum                & Model accuracy                                                  & $\checkmark$   \\
        Ma \cite{ma2022federated}             & Financial incentive     & Model accuracy                                                  & $\checkmark$   \\
        Qu \cite{qu2020decentralized}    & Not mentioned           & Data size                                                       & -              \\
        Qu \cite{qu2022fedtwin}          & Not mentioned           & Computing power, local data                                     & -              \\
        Rehman \cite{Rehman2021trustfed} & Token                   & Reputation                                                      & $\checkmark$   \\
        Wang \cite{wang2022blockchain}   & Reputation, revenue     & Reputation, shaply values, and model aggregration               & $\checkmark$   \\
        Weng \cite{weng2021deepchain}    & Token                   & Model accuracy, block mining                                    & -              \\
        Xu \cite{xu2021bafl}             & Token                   & Model accuracy, training time                                   & $\checkmark$   \\
        Xu \cite{xu2022mudfl}            & Not mentioned           & Model accuracy, consensus Participation                         & -              \\
        Zhang \cite{zhang2022blockchain} & Token                   & Training speed, computing power, and feature extractors sharing & $\checkmark$   \\
        Zhang \cite{zhang2021refiner}    & Ethereum                & Model accuracy, data size                                       & $\checkmark$   \\
        Wang \cite{wang2023incentive}    & Token                   & Model accuracy, block mining                                     & -   \\
        He \cite{he2023game}    & Token                   & Model accuracy, node behavior                                     & $\checkmark$   \\
        Wahrstatter \cite{wahrstatter2024openfl}    & Ethereum     & Model accuracy, reputation                                     & $\checkmark$   \\
        \bottomrule
    \end{tabular}
\end{table*}

The incentive basis refers to the criteria that the system uses to reward clients, which may include factors such as node reputation, data quality and quantity, and learning behavior.
For instance, Qu \emph{et al.} rewarded the clients based on the amount of data they contributed \cite{qu2020decentralized}, but this approach may not accurately reflect the overall contribution of a client to the global model.
Factors such as data quality and participation frequency can also significantly impact the effectiveness of the training process.
In contrast, Li \emph{et al.} focused solely on model accuracy as the basis for awarding nodes, as it is verifiable and reflects their contribution \cite{li2021byzantine}.
Meanwhile, Gao \emph{et al.} argued that rewards should be based on both model accuracy and node reputation, as this incentivizes continued contributions to the global model \cite{gao2022fgfl}.
In addition, to compensate the data owner, Zhang \emph{et al.} considered the energy consumption of the data owner during training and incorporated this factor into the calculation of rewards \cite{zhang2022blockchain}.

Incentives refer to the rewards that clients receive in a system, and they can take various forms such as economic items, tokens, and reputation.
Economic items provide monetary benefits to data owners, such as cryptocurrencies like Bitcoin or Ethereum.
Tokens, on the other hand, are generated by the BC-FL system and can be used to purchase services within the system, including trained models or tasks for model training.
The circulation of tokens promotes a self-sustaining ecosystem within the system that encourages participants to contribute and collaborate.
In  \cite{chen2022dimds,gao2022fgfl,weng2021deepchain}, researchers have utilized tokens within their proposed BC-FL systems as rewards
Liu \emph{et al.} used Ethereum as a reward for training, providing real-world economic incentives \cite{liu2020secure_5g}.
In addition to cryptocurrency rewards, Abdel \emph{et al.} proposed a BC-FL system for the Industrial Internet of Things that offers clients maintenance services or discounts on products from manufacturers as incentives \cite{abdel2022privacy}.

The incentive algorithm determines the specific implementation method of the incentive mechanism.
Generally, the algorithm involves quantifying each incentive basis and inputting it as a variable into the reward function, which yields the corresponding reward value.
For instance, xu \emph{et al.} devised a rewarding formula that takes model accuracy and training time into account \cite{xu2021bafl}.
The reward $cs_i$ of the $i$-th client in the proposed solution is calculated as:
\begin{equation}
    cs_i=\frac{\sum_{j=1}^{n}[\alpha\times  (acc^j_i-aggAcc^{j-1})+\frac{1-\alpha}{timeE^j_i-timeS_i^j}]}{n},
\end{equation}
where $n$ denotes the number of training rounds in which the $i$-th client participated, $acc_i^j$ denotes the model accuracy of client $i$ in round $j$, and $aggAcc^{j-1}$ represents the global model accuracy in round $j-1$.
Additionally, $timeE_i^j$ and $times_i^j$ indicate the end time and start time of the jth round for client $i$.
Furthermore, the introduction of variable $\alpha$ allows for adjustment according to different FL tasks.
If the task is more sensitive to time, the value of $\alpha$ can be reduced, while if the task is more sensitive to accuracy, the value of $\alpha$ can be increased.

\subsection{Security Enhancement}

\begin{figure}
    \setlength{\abovecaptionskip}{0pt}
    \setlength{\belowcaptionskip}{0pt}
    \centering
    \includegraphics[width=4.5in]{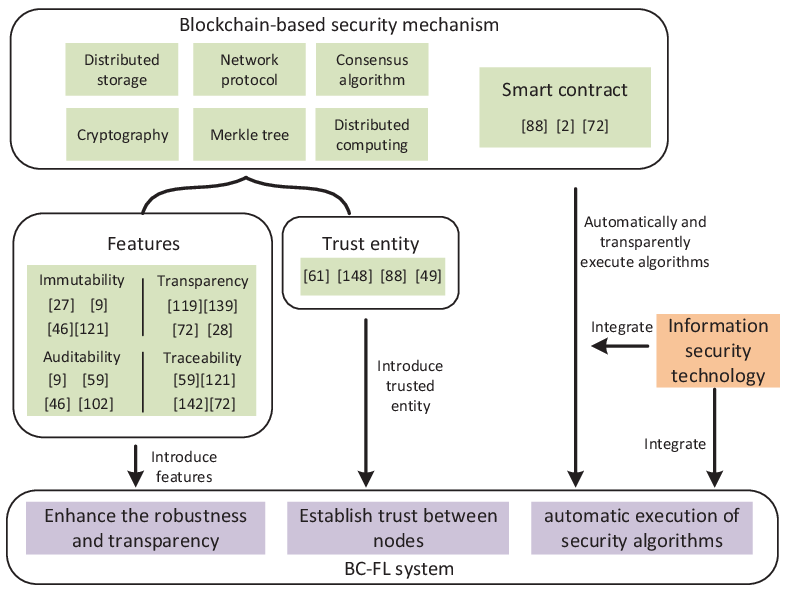}
    \caption{Security provided by blockchain for the BC-FL system.
        By employing appropriate techniques, blockchain can impart its security features  (e.g. immutability and traceability) to the FL system.
        Moreover, in a partially trusted FL environment, blockchain can act as a reliable entity to foster trust relationships.
        Furthermore, deploying security-enhancing algorithms on the blockchain via smart contracts can further enhance the security of the BC-FL system.}
    \label{fig_safty}
    \vspace{-15pt}
\end{figure}

The BC-FL system achieves the establishment of a trustworthy relationship in the system through blockchain technology.
As a distributed database, blockchain aligns with the distributed nature of FL.
With certain consensus mechanisms, the blockchain can still maintain the consistency and correctness of the system even in the presence of malicious clients.
Therefore, the robustness of blockchain against malicious nodes makes it well-suited for an environment where malicious nodes could exist in the FL system.
Furthermore, due to the robustness of the blockchain, the BC-FL system allows for the storage of vulnerable data in the blockchain, enhancing the security of the entire system.
The security issues in the BC-FL system is illustrated in Fig. \ref{fig_safty}.

To explicate the specific security properties of blockchain necessary for implementation in a BC-FL system, we conducted an extensive study of representative BC-FL systems from recent years.
The results of this research are presented in Table \ref{table:2}.


\textbf{Transparency:} Transparency is one of the key features of the blockchain.
All the information stored on the blockchain is accessible to full nodes, while light nodes can query certain information by sending requests to the full nodes.
In the BC-FL system, transparency refers to the transparent operation of algorithms and the disclosure of data.
This includes but is not limited to, the parameter aggregation operation, the reputation of each node, and the reward operation of the system.
The transparent nature of blockchain is derived from the distributed maintenance of the blockchain across all nodes in the network, with each node maintaining a local copy of the blockchain ledger.

\textbf{Auditability:} Auditability is a significant feature of blockchain systems, enabling the tracing and analysis of data using specialized algorithms.
In the BC-FL system, auditability becomes particularly valuable when specific circumstances arise, such as ineffective model training or the need to review client operations.
The recorded data on the blockchain - including local gradients - can be extracted for detailed analysis.
By analyzing previously recorded information on the blockchain, such as local gradients, nodes can be penalized for producing undesirable outcomes.

\textbf{Anti-malicious nodes:} In blockchain systems, malicious nodes can take on various forms, including those that propagate false blocks or launch attacks against the system.
Byzantine robust consensus algorithms can be used to mitigate these types of malicious behavior.
In the BC-FL system, malicious nodes are those that can undermine the effectiveness of the system, such as through poisoning attacks or privacy violations.
To address these issues, specific consensus algorithms can be designed to thwart malicious activity, or security techniques can be incorporated into the system via smart contracts.
Anti-malicious nodes and auditability both play a role in dealing with malicious nodes, but the former aims to prevent the impact of malicious nodes in real time, while the latter focuses on identifying the source of the attack after the fact.

\textbf{Traceability:} The blockchain system inherently preserves all state changes since its genesis block.
When tracing back to a previous state, the system can be readily restored to a specific point in history.
In the BC-FL system, traceability refers to the ability to restore a previously trained model or parameters saved by the current work in case of severe damage or loss due to central server failure.

\textbf{Immutability:} The immutable nature of the blockchain can be attributed to the sound design that underlies its consensus algorithm.
Each full node in a blockchain network maintains a local copy of the ledger, which ensures that malicious nodes are unable to dictate terms to other nodes unless they comply with the consensus algorithm.
Any attempts to tamper with the local copy by modifying incorrect blockchains will result in the creation of new blocks that cannot be recognized by other honest nodes.
Therefore, as long as the majority of computing power is held by honest nodes, the blockchain remains immutable.
In the BC-FL system, critical information such as client reputation and model hash values can be securely stored on the blockchain to ensure the accuracy of this data.

\textbf{Anti-single point of failure:} The term "single point of failure" refers to a scenario where a sole parameter server becomes the bottleneck for FL security, rendering the entire system inoperable if it fails due to an attack or power outage, among other reasons.
To tackle this problem, the BC-FL system replaces the role of the parameter server with blockchain technology.
As discussed in Section \ref{bc_fl_empower}, the issue of single point of a failure is elaborated upon.

\begin{table*}
    \setlength{\abovecaptionskip}{0pt}
    \setlength{\belowcaptionskip}{0pt}
    \caption{Statistics on the Security Purpose of Introducing Blockchain in BC-FL Systems.}
    \label{table:2}
    \scriptsize
    \renewcommand{\arraystretch}{1.2}
    \centering
    \tabcolsep2pt
    \begin{tabular}{L{0.8in}  C{0.65in}  C{0.65in}  C{0.8in} C{0.65in} C{0.65in} C{0.75in}}
        \toprule
        Ref.                                   & Transparency & Auditability & Anti-malicious nodes & Traceability & Immutability & Anti-single point of failure \\
        \midrule
        Awan \cite{awan2019poster}             & -            & $\checkmark$ & $\checkmark$         & -            & $\checkmark$ & -                            \\
        Cheng \cite{cheng2021blockchain}       & -            & -            & $\checkmark$         & -            & -            & -                            \\
        Feng \cite{feng2021two}                & -            & -            & $\checkmark$         & -            & $\checkmark$ & -                            \\
        Jia \cite{jia2021blockchain}           & -            & $\checkmark$ & -                    & -            & $\checkmark$ & -                            \\
        Li \cite{li2021byzantine}              & -            & $\checkmark$ & $\checkmark$         & $\checkmark$ & -            & $\checkmark$                 \\
        Liu \cite{liu2020secure}               & -            & -            & $\checkmark$         & -            & -            & -                            \\
        Lu \cite{lu2020blockchain_5G}          & -            & -            & -                    & -            & -            & $\checkmark$                 \\
        Miao \cite{miao2022privacy}            & $\checkmark$ & -            & $\checkmark$         & $\checkmark$ & -            & $\checkmark$                 \\
        Mothukuri \cite{mothukuri2021fabricfl} & -            & $\checkmark$ & -                    & -            & $\checkmark$ & -                            \\
        Sun \cite{sun2021permissioned}         & -            & $\checkmark$ & $\checkmark$         & -            & $\checkmark$ & -                            \\
        Xu \cite{xu2021bafl}                   & $\checkmark$ & -            & -                    & -            & $\checkmark$ & $\checkmark$                 \\
        Zhang \cite{zhang2022blockchain}       & $\checkmark$ & -            & -                    & -            & $\checkmark$ & -                            \\
        Zhang \cite{zhang2021bc}               & -            & -            & -                    & $\checkmark$ & $\checkmark$ & $\checkmark$                 \\
        Huang \cite{huang2023distance}         & -            & -            & -                    & -            & $\checkmark$ & -                             \\
        Ouyang \cite{ouyang2023artificial}     & -            & $\checkmark$ & $\checkmark$         & -            & $\checkmark$ & $\checkmark$              \\
        Lin \cite{lin2023drl}      & -            & - & $\checkmark$         & -            & -  & $\checkmark$              \\
        Aloqaily \cite{aloqaily2023reinforcing}    & -            & -   & $\checkmark$         & -            & -  &     -              \\
        Mu \cite{mu2023digital}    & -            & -   & $\checkmark$         & -            & -  &     $\checkmark$              \\
        Fu \cite{fu2023incentive}    & $\checkmark$            & -   & -         & -            & $\checkmark$  &     $\checkmark$              \\
        He \cite{he2023game}    & -            & $\checkmark$   & $\checkmark$         & -            & $\checkmark$  &     $\checkmark$              \\
        Xu \cite{xu2023blockchain}    & -            & -   & $\checkmark$         & $\checkmark$            & $\checkmark$  &     -              \\
        \bottomrule 
    \end{tabular}
    \vspace{-15pt}
\end{table*}

To provide a comprehensive understanding of the utilization of blockchain technology in enhancing the security of the BC-FL system, we will discuss prominent literature in this field.

In  \cite{awan2019poster}, Sana \emph{et al.} regarded the blockchain as an immutable, decentralized, and reliable entity, which they incorporate into their proposed BC-FL framework called blockchain-based privacy-preserving federated learning  (BC-based PPFL).
The utilization of blockchain provides auditability, thereby enhancing the resilience of BC-based PPFL against malicious clients.
Specifically, the assumption of semi-honest clients in the universal FL system is further elevated to the assumption of malicious clients.
To augment the credibility and dependability of the FL system, Qi \emph{et al.} introduced the adoption of smart contracts to handle FL tasks \cite{qi2022high}. These smart contracts encompass various functions such as task initiation, member selection, federated learning execution, reputation evaluation, reward distribution, and query processing.
In  \cite{zhao2020privacy}, Zhao \emph{et al.} combined Multi-Krum with reputation mechanisms as well as aggregation mechanisms to rule out malicious gradients and penalize malicious clients.
In  \cite{abdel2022privacy}, Qi \emph{et al.} proposed a smart contract called Hunter Contract  (HC) to prevent malicious clients.
HC acts as a hunter by randomly selecting a client and verifying whether the gradient uploaded by that client causes a decline in the global model accuracy.
If the reduction surpasses a predefined threshold, the client is classified as malicious.
In a blockchain system, individual nodes follow the consensus mechanism to ensure the consistency, validity, and accuracy of the data.
In a BC-FL system, the data or training results of the FL process are stored on the blockchain, and the blockchain's consensus mechanism can be used to verify the content of the FL.
Consequently, some researchers have improved the security of FL by adjusting the blockchain's consensus mechanism.

In  \cite{li2021byzantine}, Li \emph{et al.} proposed a Byzantine-resistant consensus mechanism named Proof of Accuracy, which serves to identify models of poor quality.
This consensus algorithm takes into consideration not only the exclusion of local models that are deemed too poor for aggregation into the global model but also the potential for a local model with a high loss value to aid the global model in escaping local optimal solutions.
To fulfil this requirement, the consensus algorithm employs two critical thresholds: the accuracy oscillation threshold  (AOT) and the accuracy deviation threshold  (ADT).
The AOT determines the maximum acceptable accuracy reduction permitted by the accepted model, while the ADT determines the maximum absolute difference in accuracy among different client models.
These two thresholds are subject to dynamic adjustments as the algorithm progresses.
In  \cite{qiu2022rendering}, Qiu \emph{et al.} increased the security of the BC-FL system through the introduction of a novel consensus protocol called Proof of Learning  (PoL).
In contrast to PoW, PoL requires nodes to compete for the privilege of accounting rights through calculation by training a FL model, where the node with the smallest loss value adds a new block as the winner.
Other clients aggregate the winner's local model based on the reputation value against the winning node after verifying the authenticity of the newly added block.
Ouyang \emph{et al.} utilized smart contracts to authenticate participating nodes and prevent malicious nodes from participating \cite{ouyang2023artificial}.

As mentioned earlier, certain security technologies from the field of information security have been considered for use in the BC-FL system to enhance their security.
While not directly related to the security of the BC-FL system, smart contracts can serve as a platform for running certain algorithms.
Hence, we will provide a brief overview of this topic.
To safeguard client privacy, the utilization of homomorphic encryption and differential privacy algorithms \cite{yin2021comprehensive} is common, and researchers have developed advanced algorithms building upon these fundamental techniques.
We have organized this material in Section \ref{secure_challenges_and_solutions}.

\section{Challenges and Solutions in BC-FL Systems}
\label{bc_fl_challenges}


While blockchain can indeed enhance the capabilities of the FL systems and mitigate certain limitations, it is imperative to duly recognize and confront the accompanying drawbacks.
In this section, we will delve into the principal challenges entailed in the integration of blockchain into FL and the corresponding solutions, which can be broadly classified into three key aspects: efficiency, security, and storage.

\subsection{Efficiency Challenges and Solutions}
\label{efficiency_challenges_and_solutions}
\begin{figure}
    \centering
    \includegraphics[width=4.5in]{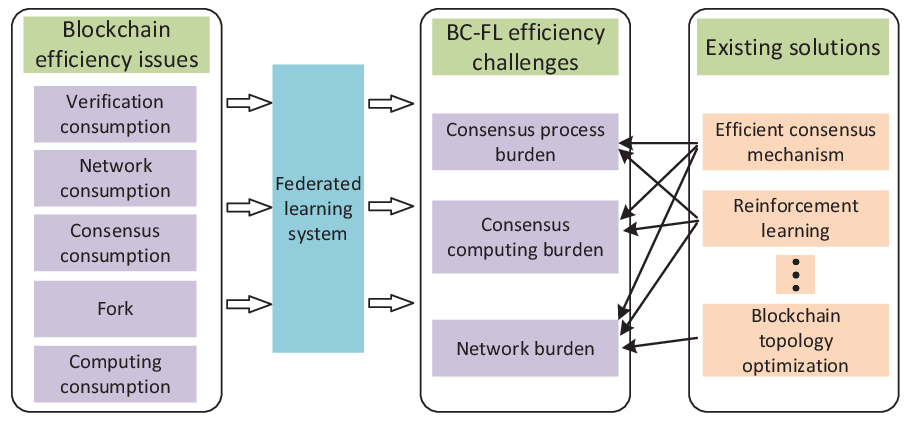}
    \caption{Efficiency challenges and related solutions in the BC-FL systems.
        The efficiency of blockchain is susceptible to factors such as network and computing overhead.
        Consequently, BC-FL systems potentially lead to a decrease in overall efficiency.
        In response to thus challenges, multiple strategies are contributed to mitigate the reduction in system efficiency.
        These methods include but not limited to, efficient consensus mechanisms, reinforcement learning, and optimized blockchain topologies.}
    \label{fig_effective}
    \vspace{-15pt}
\end{figure}

The processing capacity of blockchain systems is inherently limited.
For instance, Bitcoin can only handle seven transactions per second \cite{bamakan2020survey}.
In contrast, modern centralized payment systems can process thousands of transactions per second \cite{fan2020performance}.
Fig. \ref{fig_effective} illustrates the efficiency challenges faced by blockchain in BC-FL systems.
Unlike centralized systems, blockchain systems necessitate additional steps such as verification, communication, and network-wide consensus to maintain normal operations, which reduces the efficiency when integrating blockchain with FL systems.
Current BC-FL systems address these efficiency issues through various methods, including efficient consensus mechanisms, reinforcement learning (RL), and optimized blockchain topologies.
A summary of the pertinent literature is provided in Table \ref{table:efficiency}.

\begin{table*}
    \setlength{\abovecaptionskip}{0pt}
    \setlength{\belowcaptionskip}{0pt}
    \caption{Comparison between different solutions for each efficiency challenge.}
    \label{table:efficiency}
    \renewcommand{\arraystretch}{1.2}
    \centering
    \scriptsize
    \tabcolsep 5pt
    \resizebox{0.8 \linewidth}{!}{
    \begin{tabular}{m{0.7in}<{\centering}  m{1.00in}<{\centering}  m{1.25in}<{\centering}  m{1.00in}<{\centering} m{1.4in}<{\centering}}
        \toprule
        Ref.                              & Solutions                       & Detailed methods                                     & Dataset                & Evaluation indicators                                       \\
        \midrule
        Cao \cite{cao2021toward}          & Blockchain topology             & DAG blockchain                                       & MNIST                  & Accuracy, loss, iteration delay                             \\
        Cheng \cite{cheng2021blockchain}  & Consensus algorithm, blockchain topology& Two-layer blockchain, Raft, PBFT                     & -                      & Latency reduction                                           \\
        Feng \cite{feng2021two}           & Consensus algorithm, blockchain topology& Two-layer blockchain, sharding                       & MNIST                  & Accuracy, time cost                                         \\
        Hieu \cite{hieu2022deep}          & RL                              & DRL                                                  & -                      & Energy consumption, latency, total payment                  \\
        Li \cite{li2020blockchain}        & Consensus algorithm             & committee consensus mechanism                        & FEMNIST                & accuracy, communication overhead                            \\
        Lu \cite{lu2020blockchain}        & RL, blockchain                  & DAG blockchain, DRL                                  & -                      & Accuracy, time cost, agent reward, cumulative cost          \\

        Nguyen \cite{nguyen2021federated} & Consensus algorithm             & Proof of reputation                                  & DarkCOVID, ChestCOVID  & Running latency, block verification latency, Accuracy, Loss \\
        Nguyen \cite{nguyen2022latency}   & RL                              & DRL, A2C                                              & SVHN, Fashion-MNIST    & Accuracy, agent reward, latency, Loss                       \\
        Qi \cite{qi2021blockchain}        & Consensus algorithm             & Modified PBFT                                        & Diabetes Breast Cancer & Accuracy, time cost, gas cost                               \\
        Qu \cite{qu2022fedtwin}           & Consensus algorithm             & Proof of federalism                                  & CIFAR-10               & Accuracy                                                    \\
        Xu \cite{xu2022mudfl}             & Consensus algorithm, blockchain topology& Two-layer blockchain, proof of credit, efficient BFT & MNIST                  & Latency, communication overhead, data throughput            \\
        Zhao \cite{zhao2021deep}          & RL                              & Federated DDQL                                       & -                      & Agent reward, latency                                       \\
        Wang \cite{huang2023distance}          & Blockchain topology      &   Two-layer blockchain                                 & TSP, FMNIST             & Accuracy, energy consumption, learning utility                                     \\
        Yuan \cite{yuan2024secure}          & Consensus algorithm, blockchain topology & Blockchain sharding, DAG-based mainchain   & MNIST,  Penn Treebank   & Accuracy, training Latency, testing perplexity   \\
        Lin \cite{lin2023drl}          & Blockchain topology, RL & Blockchain sharding, DRL-based sharding & MNIST, KMNIST, FMNIST,  CIFAR-10 & Accuracy, agent reward,  reputation of nodes   \\
        \bottomrule
    \end{tabular}
    }
    \vspace{-15pt}
\end{table*}

\subsubsection{Efficient Consensus Algorithms}
The PoW consensus protocol provides robust resistance against Sybil attacks on the public chain, ensuring a strong defense against malicious nodes.
However, a primary drawback of the PoW mechanism lies in its requirement for a block generation rate that is slower than the rate of block propagation across the network, aimed at minimizing the risk of a blockchain fork.
Current research reveals relatively modest transactions per second (TPS) for both PoW and PoS consensus protocols \cite{mingxiao2017review}.
Typically, the PoW protocol achieves TPS figures below 100, while the PoS protocol reaches less than 1000 TPS.
In actuality, the TPS tends to be even lower; for instance, Bitcoin operates at a mere 7 TPS \cite{gobel2017increased}.
Additionally, the competitive nature among miners vying for block mining rewards escalates energy consumption.
Moreover, suboptimal network conditions of edge devices heighten the likelihood of forks.
In the BC-FL systems, underpinned by a partially decentralized architecture, the scenario improves to some extent.
Nonetheless, achieving consensus across the entire network still requires considerable time, impeding the speed of model aggregation.
Consequently, numerous researchers are dedicating their efforts to the development of efficient consensus protocols that can enhance the overall operational velocity of the BC-FL system, all the while aligning with the requirements of the federated learning process.


In \cite{nguyen2021federated, xu2019improvement}, Nguyen and Xu \emph{et al}. contended that certain established consensus algorithms introduce substantial communication overhead while striving for consensus.
For example, DPoS necessitates that each blockchain node communicates with a minimum of half the nodes within the BC-FL system for confirmation, leading to redundant validations among these nodes.
To tackle this challenge, they designed a streamlined consensus mechanism known as Proof of Reputation (PoR).
Within the POR algorithm, every blockchain node is permitted to validate with just a single other node during the consensus process, resulting in a significant reduction in validation delays.
In \cite{qu2022fedtwin}, Qu \emph{et al.} introduces a Proof-of-Federalism (PoF) consensus algorithm, which builds upon the foundation of PoW.
PoF leverages the training of FL tasks as a viable alternative to the challenge of discovering a fitting nonce in PoW, effectively sidestepping the computational resources typically expended during the consensus calculation process.
Before each training round commences, intelligent contracts sift through unfavorable local model parameters and cherry-pick local models that lend themselves well to global aggregation.
During cross validation, each node singles out the most optimal set of global models.
Upon reaching a predetermined time threshold, the participant who boasts the highest number of selected global models emerges as the victorious contender.

In \cite{xu2022mudfl}, Xu \emph{et al.} proposed a lightweight blockchain network for FL systems called  micro-chain to address the issues of low transaction throughput and poor scalability.
Participants in FL are divided into multiple small-scale micro chains, each of which is unified through an advanced inter-chain network using Byzantine fault-tolerant consensus protocols.
Within each micro-chain, block consistency is achieved using the Proof of Credit (PoC) algorithm, where committee members are responsible for generating new blocks.
Then, a new committee is randomly selected at the end of each dynasty round.
Ledger consensus is achieved using the Vote-based Chain Finality (VCF) protocol, where committee member nodes vote to select the preferred branch in case of network forks.
In \cite{li2020blockchain}, Li \emph{et al.} introduced an innovative committee consensus mechanism aimed at significantly reducing the required consensus computation.
The proposed mechanism selects multiple clients as committee nodes in each training round, utilizing the data on these committee nodes as the validation set.
The final scores for each trained client are then determined by taking the median of the scores of these clients.
These scores are subsequently used to perform global model aggregation by selecting a specific number of clients with the highest scores.



\subsubsection{Reinforcement Learning}
RL is a machine learning algorithm that enables an agent to interact with the environment, learn from its experiences, and take action accordingly.
The ultimate objective is to maximize the cumulative reward obtained by the agent over time.
The traditional optimization methods are ineffective in the BC-FL system because of the system's complexity, a large number of participants, and their limited computing and communication resources.
To address these challenges and achieve better results, RL can be utilized to optimize resource allocation and schedule the resources of each client based on signals received from them.
This can potentially reduce system delays and lead to improved performance.

To apply RL in the BC-FL system, there are several fundamental steps to follow.
First, the system designer must define the environment based on specific circumstances, such as the parameters of the client and network conditions.
This environment can be modelled as a Markov decision process.
Second, the agent's action space should be defined, which includes factors such as the energy consumed by the device during training and the block generation difficulty.
Third, defining the reward is essential.
In general, the reward in the system can be based on overall training delay that encourages the agent to find ways to reduce the system delay effectively.
Finally, RL training is performed using a specific algorithm.
The agent learns how to optimize resource allocation within the BC-FL system under different environmental scenarios through continuous interaction with the environment.

In  \cite{hieu2022deep}, Hieu \emph{et al.} used the deep reinforcement learning method \cite{arulkumaran2017deep} to control the data and energy used for training and block generation in the device.
By judiciously allocating resources, they were able to mitigate the system delay and enhance overall system efficiency.
In  \cite{lu2020blockchain}, Lu \emph{et al.} used the Deep Q-learning (DQL) \cite{arulkumaran2017deep} method to facilitate client selection for the FL process.
They formulated a joint optimization plan by considering the client's available wireless transmission rate, client computing power  (CPU frequency), and the current selection status of clients as the state of the DQL method.
The reward function is designed as a weighted sum of the loss function of each node, the computation time, and the communication time.
This approach leads to a high level of model accuracy while maintaining a low global system cost.
The proposed algorithm design shows promising results in performance evaluation, indicating its potential in real-world applications.

In  \cite{zhao2021deep}, Zhao \emph{et al.} proposed a BC-FL system for vehicle networks.
The proposed system allows autonomous vehicles  (AVs) to offload part of their computing tasks to edge servers  (ESs), effectively reducing local computation latency, communication latency, and blockchain consensus latency.
To achieve this, the authors employed a federated duel deep Q-learning  (DDQL) algorithm \cite{wang2016dueling} and deployed it to each AV to enable them to take action according to the changing external environment.
The state space of the proposed DDQL includes wireless channel conditions, data set quality, and packet error rate, where AVs select offload strategy, wireless channel, and CPU-cycle frequency based on the DDQL algorithm.

In  \cite{nguyen2022latency}, Nguyen \emph{et al.} applied the DRL method based on a parameterized advantage actor-critic  (A2C) algorithm \cite{zahavy2020self} to a multi-server edge computing scenario to reduce the overall system latency.
Their proposed hybrid discrete-continuous action DRL algorithm takes into account various factors such as data size, channel state, broadband state, computation state, and hash power to determine whether an edge node should perform computation offloading.
In case of offloading, the agent needs to decide on the corresponding channel selection, power allocation and other transmission necessary parameters.
In case of non-offloading, the agent needs to decide on the necessary parameters for training such as the hash power allocation for local computation.
Unlike existing purely discrete or purely continuous action DRL algorithms, the authors proposed a hybrid model where resource allocation is continuous, while the offloading decision is discrete, leading to improved training performance.


\subsubsection{Optimized Blockchain Topology}
The topological structure of a blockchain system is a crucial factor that impacts information transmission and significantly influences the system's efficiency and scalability.
Modifying the topology of the blockchain can potentially improve its efficiency, which has been demonstrated in some papers in the BC-FL systems \cite{hao2020towards,xie2019survey}.
The topology of a blockchain system includes the physical and logical topology, both of which can affect the system's efficiency.

Improving the physical topology involves considering the node layout, physical location, and network topology.
For instance, positioning relevant nodes near the data source can reduce the network delay, altering node connections' topology can enhance network transmission efficacy, and using edge computing can reduce the computing burden of edge devices.

The logical topology of the blockchain refers to how transactions and blocks are verified and added, and it can impact the processing speed and scalability of the system.
Compared to traditional blockchains, the DAG blockchain \cite{cui2019efficient} is better suited for scenarios such as FL that require efficient processing of large amounts of data due to its faster transaction processing speed and better scalability.
The primary difference between the DAG blockchain and the traditional blockchain is the way data is organized and stored.
The latter uses chain storage, where a block can point to at most one previous block and forks are not allowed.
On the other hand, the DAG blockchain employs directed acyclic graph organization, allowing multiple previous blocks to be pointed to as long as no cycle is formed.
This data organization form enables the block generation speed to exceed the block propagation speed, resulting in high concurrency and weak synchronization.

In addition, there are some improvements that involve both the physical and logical topology of the blockchain.
One such improvement is the deployment of a two-layered blockchain architecture, which comprises two relatively autonomous blockchains – the main-chain and the sub-chain.
The sub-chain is responsible for interfacing with peripheral devices and executing swift consensus algorithms.
Meanwhile, a subset of nodes within the sub-chain are nominated to constitute the main-chain.
Typically, the main-chain utilizes Byzantine fault-tolerant consensus algorithms to ensure the security of the system.

In  \cite{cheng2021blockchain}, Cheng \emph{et al.} proposed a BC-FL system based on a two-layer blockchain architecture.
The lower-layer blockchain is responsible for connecting devices to achieve strong consistency and a high consensus rate.
Within a short period of time, the lower-layer blockchain needs to reach a consensus while only considering the problems of equipment failure and omission.
To this end, the Raft protocol is employed, which is more efficient despite lacking Byzantine fault tolerance.
The upper-layer blockchain connects various lower-layer blockchains and is designed to prevent malicious nodes and resolve Byzantine faults.
Thus, the PBFT algorithm is employed, which can effectively resist Byzantine attacks but requires a longer time frame for consensus.
The upper-layer blockchain's nodes are super nodes with robust computing power selected from the lower-layer blockchain.
Hence, they are relatively small in number but possess significant computing capabilities, enhancing the PBFT protocol's consensus speed.

\subsection{Secure Challenges and Solutions}
\label{secure_challenges_and_solutions}
\begin{figure*}
    \centering
    \includegraphics[width=4.5in]{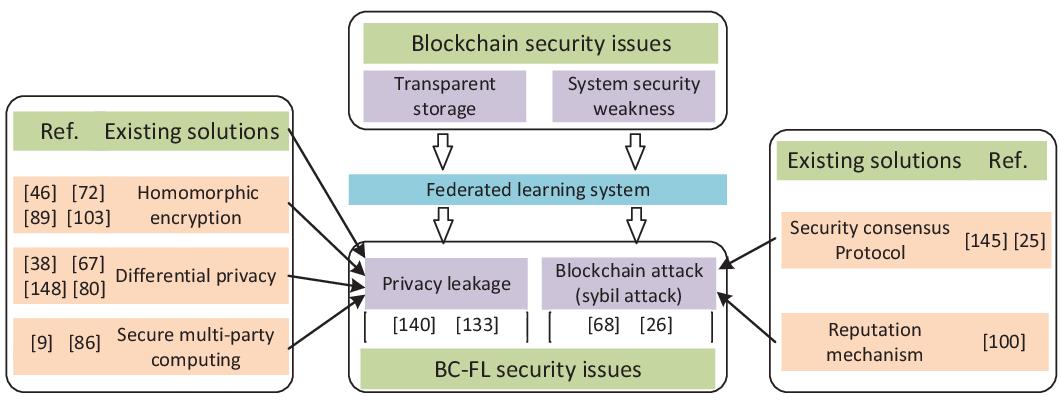}
    \caption{Security challenges and related solutions in the BC-FL systems.
        Malicious nodes pose a threat to the blockchain within the BC-FL system through two distinct avenues: privacy leakage and consensus mechanisms.
        The former capitalizes on the blockchain's data transparency to breach access to model information stored within it, whereas the latter employs attacks via the straightforward consensus mechanism inherent in the BC-FL system.
        In response to these challenges, contemporary solutions are predominantly centered around the development of diverse privacy protection algorithms and the implementation of exceptionally secure consensus mechanisms.
    }
    \label{fig_insecurity}
    \vspace{-15pt}
\end{figure*}

Integrating blockchain into FL systems holds the potential to significantly bolster system security.
However, the successful execution of such integration in BC-FL systems hinges greatly upon the scrupulous deliberation of system designers and the implementation of effective combination strategies.
Inadequate integration of blockchain may give rise to supplementary predicaments.
The security challenges and related solutions in the BC-FL systems are evidenced in Fig. \ref{fig_insecurity}.

As shown in Fig. \ref{fig_insecurity}, the transparent nature of blockchain data raises concerns about storing sensitive information, potentially leading to violations of privacy.
Additionally, extant attack methods targeting blockchain systems, such as Sybil attacks \cite{zhang2019double}, have the capability to compromise the security of the BC-FL system.
An examination of recent BC-FL systems has unveiled several instances wherein Sybil attacks and breaches of privacy remain plausible.

\subsubsection{Privacy Leakage}
The immutability and transparency inherent in blockchain play a pivotal role in safeguarding the integrity of a system.
Blockchain data can be validated by all clients, and it remains impervious to unauthorized tampering by malicious entities.
However, this approach also brings forth a potential vulnerability, as malevolent nodes can effortlessly access sensitive data stored on the blockchain.
In BC-FL systems, multiple research endeavors permit clients to store local models or gradients on the blockchain, along with their retrieval methods \cite{zhang2022privacy,yuan2022secretgen}.
Regrettably, this allowance opens the door for malicious clients to potentially deduce sensitive worker data.
To tackle this issue, several scholars suggested the implementation of diverse cryptographic techniques \cite{feng2021blockchain, majeed2021st}.
These techniques serve to fortify the system's privacy protection capabilities while mitigating the potential privacy hazards.

Homomorphic Encryption (HE) represents an encryption technology that facilitates direct computations on encrypted data, empowering aggregators to execute model aggregation operations without necessitating the decryption of local models \cite{qi2021blockchain, sun2022blockchain, jia2021blockchain, miao2022privacy}.
In \cite{sun2022blockchain}, Sun \emph{et al}. designed a Bresson-Catalano-Pointcheval based homomorphic noise mechanism to secure gradient values and pinpoint malevolent data owners.
Meanwhile, in \cite{jia2021blockchain}, Jia \emph{et al.} seamlessly incorporated the homomorphic encryption scheme Paillier \cite{acar2018survey} into K-means clustering, distributed random forest, and distributed AdaBoost components in the BC-FL systems.
The scheme offers a privacy-preserving solution for client data when employing these machine learning algorithms.
In another study \cite{miao2022privacy}, Miao \emph{et al}. harnessed Fully Homomorphic Encryption (FHE) to facilitate secure model aggregation.
Concurrently, they leveraged blockchain to ensure the transparency of the aggregation process.
In \cite{sun2021permissioned}, Sun \emph{et al.} introduced a verification procedure in [104] before local update aggregation to fend off poisoning attacks. They introduced differential privacy noise during the verification process to obfuscate local updates, thereby enhancing privacy.
Additionally, in \cite{fang2022privacy}, Fang \emph{et al.} outlined a secure and verifiable local update aggregation scheme, replacing differential privacy technology with the Shamir Secret Sharing technique\cite{li2019privacy} to ensure the correctness of confidential sharing.

Multiple studies also employed differential privacy to protect the privacy of FL clients \cite{ma2022federated,nguyen2021federated,he2022cgan}.
In \cite{ma2022federated}, Ma \emph{et al}. delved into a differential privacy solution for the BC-FL system, where noise is added to the local data features to uphold local privacy and pseudo-noise sequences are adopted to identify inactive clients.
Similarly, in \cite{abadi2016deep}, Abadi \emph{et al}. incorporated tailored noise into the data prior to sharing, effectively obscuring the actual data values while maintaining usability even after noise integration.
Within BC-FL systems integrating differential privacy, it is customary for clients to introduce noise to the model prior to uploading the local model, thereby ensuring privacy protection.
In \cite{zhao2020privacy}, Zhao \emph{et al.}  employed differential privacy to safeguard the privacy of individual clients by applying it to the extracted data features of each client. Additionally, Qu's work \cite{qu2022fedtwin} presented an enhanced differential privacy algorithm built upon generative adversarial networks, offering a means of preserving the privacy of local models.


Secure Multi-Party Computing (SMPC) stands out as another promising avenue for ensuring privacy of BC-FL systems \cite{awan2019poster, passerat2020blockchain}.
SMPC represents a versatile cryptographic tool that empowers distributed parties to collaboratively compute diverse functions while withholding their confidential inputs and outputs \cite{zhao2019secure}.
Within the BC-FL system incorporating SMPC, every client employs the SMPC protocol to join forces and aggregate the global model.
SMPC can be instantiated as a smart contract on the blockchain, with these contracts delineating computation rules and guaranteeing proper protocol execution.
In \cite{awan2019poster}, Awan \emph{et al.} designed a meticulously algorithm that leverages homomorphic encryption and proxy re-encryption grounded in the Paillier encryption algorithm.
This technique involves encrypting each local model, thereby preventing the model aggregator from accessing individual models.
Nevertheless, upon aggregating the encrypted local models, the aggregator can obtain an unencrypted global model, thus preserving the confidentiality of each client's data.


Several studies explored alternative approaches to address the privacy concerns within the BC-FL system \cite{fang2022privacy, Wei2022Redactable, guo2021lightfed}.
For instance, in \cite{Wei2022Redactable}, Wei \emph{et al.} introduced a chameleon hash scheme with a modifiable trapdoor (CHCT) as a countermeasure to potential privacy leaks on the blockchain, effectively creating an adaptable blockchain structure. The CHCT employs trapdoors to generate hash collisions, resulting in identical hash values.
When sensitive or erroneous data is identified on the blockchain, clients can utilize CHCT to amend the relevant data.
However, strict adherence to a well-defined set of procedures is imperative when modifying the blockchain to safeguard its reputation as a trusted third-party entity.
In \cite{fang2022privacy}, Fang \emph{et al}. employed a privacy-preserving strategy to store the gradient's commitment on the blockchain and mapped it to an elliptic curve point.
Simultaneously, the gradient is obscured using a Pseudorandom generator-based mask, which can subsequently be removed to restore the accurate global gradient once all local gradients are incorporated.
Similarly, \cite{guo2021lightfed}, Guo \emph{et al.} presented a blockchain-based obfuscation transmission mechanism, shielding the local models of FL edge nodes from external scrutiny by potential attack devices.
The blockchain is initially divided into distinct branches starting from the genesis block, each corresponding to a training device. A hash key block on each branch stores the hash key function published by the server.


\subsubsection{Sybil Attacks}
Sybil attacks have garnered extensive attention within the blockchain field, owing to their potential to compromise the integrity and security of blockchains \cite{zhang2019double}.
Thus attacks involve an assailant generating numerous false identities or nodes within the network, affording them the means to manipulate the system's dynamics \cite{otte2020trustchain}.
Established methods like Proof of Work (PoW) and Proof of Stake (PoS) have demonstrated some degree of resilience against Sybil attacks \cite{baza2020detecting,matzutt2020utilizing}.
Within the context of the BC-FL system, certain endeavors have adopted lightweight consensus protocols or rapid information transmission methods to bolster system speed, inadvertently rendering them susceptible to Sybil attacks \cite{lu2020blockchain, feng2021two, shayan2020biscotti}.
For instance, in \cite{lu2020blockchain}, the Raft protocol is harnessed to expedite consensus within the underlying blockchain.
However, this approach exposes a vulnerability where an attacker could subvert the leader election process through the creation of fabricated identities.
This disruption might impede the proper selection of legitimate leaders or lead the system astray from its intended behavior.
In another instance, Feng \emph{et al}. employed a localized model update chain facilitated by inter-device communication for efficient blockchain information transfer \cite{feng2021two}.
While inter-device communication offers improved network performance and reduced communication costs, it also presents a vulnerability to Sybil attacks \cite{shayan2020biscotti}.
In the realm of inter-device communication, attackers exploit the creation of multiple spurious identities or devices to gain a foothold in the network, inundating it with counterfeit traffic or acquiring sensitive information.


Another group of research tried to employ various consensus mechanisms to counter Sybil attacks \cite{zhang2021refiner,gilad2017algorand, fang2022privacy, shayan2020biscotti}.
For instance, in \cite{zhang2021refiner}, Zhang \emph{et al}. utilize a validator committee selection scheme akin to the Algorand consensus algorithm \cite{gilad2017algorand}, utilizing verifiable random numbers to thwart Sybil attacks.
In \cite{fang2022privacy}, Fang \emph{et al}. designed a secure aggregation protocol that directly applies the Algorand consensus algorithm to fend off Sybil and tampering attacks.
The protocol uses pairwise random masks to impede Sybil attacks.
Shayan \emph{et al}. \cite{shayan2020biscotti} introduced a fully decentralized system to effectively mitigate Sybil attacks by judiciously defining reputation levels.
They used blockchain and cryptographic primitives to defends against known attacks.

\subsection{Storage Challenges and Solutions}
\label{storage_challenges_and_solutions}
The storage requirements for blockchain systems are inherently cumbersome, as each full node is required to maintain a complete backup of the entire system.
This leads to a linear increase in the total storage size with the number of full nodes.
In FL, clients transmit their local models to a central server and download the global model.
The server is responsible for storing both the local and global models and various FL-related data and parameters, thereby becoming the node with the highest storage demand in the FL system.
When enhancing the FL system with blockchain, the BC-FL system must inevitably store diverse information on the blockchain, resulting in significant storage overhead.
Furthermore, most blockchain platforms currently impose limitations on transaction or block size.
For instance, Bitcoin has a block size limit of 1MB, and while Ethereum does not have a theoretical block size limit, its gas limit effectively restricts the size of transactions \cite{nagayama2020identifying}.
If the BC-FL system requires direct storage of large volumes of data within blocks, such as model parameters, this could surpass the blockchain system's storage capabilities.

\begin{figure}
    \setlength{\abovecaptionskip}{0pt}
    \setlength{\belowcaptionskip}{0pt}
    \centering
    \includegraphics[width=4.0in]{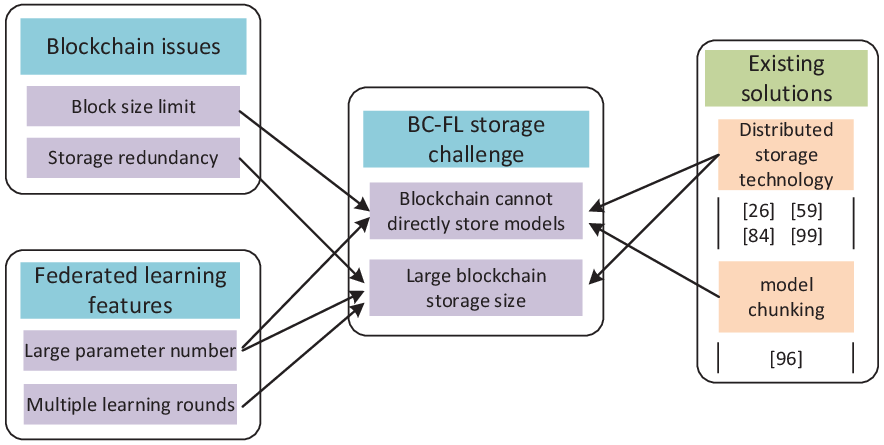}
    \caption{Storage challenges and related solutions in BC-FL systems.}
    \label{fig_storage}
    \vspace{-15pt}
\end{figure}

As illustrated in Fig. \ref{fig_storage}, the storage challenges of the BC-FL system are primarily twofold:

\textbf{Constrained storage capacity:} the limited block size makes storing some data that takes up storage space difficult.

\textbf{Redundant storage demands:} a large amount of training-related data is stored in the blockchain, which brings unnecessary information redundancy and terrible storage challenges to the entire BC-FL system.

As depicted in Fig. \ref{fig_storage}, the current landscape presents two prevailing strategies to tackle the storage challenges in BC-FL systems.
The first approach entails chunking the FL models or data into distinct segments, which are then stored on the blockchain with constrained block size \cite{ramanan2020baffle}.
This methodology necessitates prior negotiation of a serialization plan among nodes.
Subsequently, each split data's size is logged as supplementary information within the transaction block.
Gradual storage of the data on the BC-FL system is accomplished through the initiation of transactions.
However, we assert that such techniques possess restricted applicability and are suitable solely for systems characterized by a few supernodes, each endowed with robust storage capabilities capable of managing storage redundancy.

The second solution involves utilizing distributed storage technology to house the model, while retaining only the acquisition method on the blockchain \cite{Rehman2021trustfed,zhao2020privacy,majeed2021st,salim2022federated,ouyang2020learning}.
For example, the InterPlanetary File System (IPFS) employs content addressing for file storage and retrieval, allowing users to access files using the hash value associated with the file \cite{feng2021blockchain}.
In this methodology, solely the hash of the respective model finds its place on the blockchain.
Additionally, Xu \emph{et al.} incorporated a model producer within the system to provide download links to other nodes \cite{li2021byzantine}.
The blockchain then retains the model hashes and corresponding download links solely as part of this innovative approach.
These approaches address the intricate interplay between blockchain and FL requirements, paving the way for more efficient and effective storage management within BC-FL systems.

\section{Future Research Directions}
\label{future_research_directions}
In this section, we delve into prospective research avenues at the intersection of blockchain and FL systems.
These encompass concepts like combination architecture, lightweight blockchain solutions, and personalized smart contracts.
The fusion of blockchain technology with FL presents an auspicious and pioneering strategy for tackling specific challenges.
Yet, despite its pragmatic significance, the current body of research in this field remains inadequate.
Having meticulously scrutinized the latest studies, we distill a collection of potential future research directions, presented herein for consideration.

\subsection{Combination Architecture}
The majority of current research endeavors have centered around the integration of blockchain into HFL systems.
An imperative exists to delve into the synergies between blockchain and VFL, as well as federated transfer learning.
VFL presents distinct data processing and training methodologies in comparison to HFL \cite{yang2019federated}.
Prior to commencing training, VFL necessitates privacy-preserving set intersection, and the regular encryption and exchange of interim training outcomes. This raises the inquiry of whether blockchain can effectively tackle the unique challenges posed by VFL and federated transfer learning.
Further exploration is warranted to ascertain the potential of blockchain in addressing these specialized concerns within the realm of VFL and federated transfer learning.

\subsection{Lightweight Blockchain Solutions}
In FL systems, particularly in cross-device FL, clients typically exhibit constrained communication and computational capacities.
Introducing blockchain on each client might further burden the communication and computational resources of edge devices.
The majority of blockchains in BC-FL systems maintain a rather general-purpose nature, with only a handful being meticulously customized for these systems. The forthcoming challenge lies in the advancement of consensus algorithms, topology structures, communication methodologies, and other enhancements aimed at enhancing the compatibility of blockchain systems with the FL framework.

\subsection{Personalized Smart Contracts}
The integration of smart contracts has substantially elevated the adaptability and scalability of BC-FL systems.
A promising avenue for future exploration involves the formulation of supplementary algorithms tailored for deployment on personalized smart contracts, aiming to enhance the efficiency, security, and flexibility of BC-FL systems \cite{wan2021smart}.
It is important to highlight that a multitude of ongoing investigations are centered around attacks and defenses in the realm of smart contract security \cite{so2020verismart, permenev2020verx}.
Thus, while the utilization of smart contracts within BC-FL systems holds promise, a prudent approach necessitates meticulous scrutiny of potential vulnerabilities, mandating their mitigation through rigorous and comprehensive research endeavors.

\section{Conclusion}
\label{conclusion}
Blockchain-empowered Federated Learning (BC-FL) has emerged as a promising realm of distributed machine learning in recent years.
This all-encompassing review delves into the potential advantages and challenges associated with the integration of blockchain into FL.
The survey highlighted numerous domains where blockchain can be harnessed to enhance security, avert single points of failure, and establish reputation and incentive mechanisms.
We elucidated how blockchain can surmount the primary challenges encountered by FL.
By considering that the amalgamation of blockchain also presents several challenges that necessitate resolution, we succinctly outlined the efficiency, storage, and security challenges that arise in BC-FL systems, and provided a comprehensive survey of prevailing solutions.
This survey furnishes a thorough and insightful analysis of the role of blockchain in the FL system.
We hold the belief that this work will expedite the exploration and advancement of related research endeavors, thus bestowing a valuable resource upon scholars and practitioners in this field.

\bibliographystyle{unsrt}
\bibliography{references}

\end{document}